\documentclass[a4paper,fleqn,usenatbib]{mnras}

\usepackage{mathptmx}
\usepackage{txfonts}

\usepackage[T1]{fontenc}
\usepackage{ae,aecompl}

\newcommand{\Msun}{$\rm M_{\sun}$\ } 
\newcommand{\Msol}{M$_{\sun}$}   
\newcommand{\mcm}{cm$^{-3}$}         
\newcommand{\mcmb}{cm$^{-3}$ }       
\newcommand{\pccm}{particles per \mcm}   
\newcommand{\pccmb}{particles per \mcmb} 
\usepackage{graphicx}	

\usepackage{ifthen}
\newboolean{hyper}         
\setboolean{hyper}{false}   
\ifthenelse{\boolean{hyper}}{
\hypersetup{
        colorlinks=true,linkcolor=blue,citecolor=blue,filecolor=blue,urlcolor=blue,
	pdfauthor={Katharina Fierlinger},
	pdftitle={PhD project, Paper I},
	pdfsubject={PhD thesis, Paper I, 1D hydrodynamics},
	pdfproducer={emacs, latex}
}
}{}

\title[Winds or SNe?]{Stellar feedback efficiencies: supernovae versus stellar winds}

\author[Fierlinger et~al.]{Katharina M.\ Fierlinger$^{1,2,3}$\thanks{E-mail: kfierlin@usm.lmu.de}, 
Andreas Burkert$^{1,3,4}$, 
Evangelia Ntormousi$^{5}$, 
\newauthor 
Peter Fierlinger$^{2,3}$, 
Marc Schartmann$^{1,4,6}$, 
Alessandro Ballone$^{1,4}$, 
\newauthor 
Martin G.\ H.\ Krause$^{1,3,4}$, 
Roland Diehl$^{3,4}$ \\ 
$^{1}${Universit{\"a}tssternwarte der Ludwig-Maximilians-Universit{\"a}t, Scheinerstra{\ss}e 1, 81679 M{\"u}nchen, Germany}\\
$^{2}${Physik-Department TU M{\"u}nchen, James-Franck-Stra{\ss}e, 85748 Garching, Germany}\\
$^{3}${Excellence-Cluster ``Origin \& Structure of the Universe'', Boltzmannstra{\ss}e 2, 85748 Garching, Germany}\\
$^{4}${Max-Planck-Institut f{\"u}r extraterrestrische Physik, Giessenbachstra{\ss}e 1, 85748 Garching, Germany}\\
$^{5}${Laboratoire AIM, Paris-Saclay, CEA/IRFU/SAp - CNRS - Universit{\'e} Paris Diderot, 91191, Gif-sur-Yvette Cedex, France}\\
$^{6}${Centre for Astrophysics and Supercomputing, Swinburne University of Technology, Hawthorn, Victoria 3122, Australia}
}

\date{Accepted 2015 November 16. Received 2015 November 16; in original form 2015 July 28}

\pubyear{2015}

\begin{document}
\label{firstpage}
\pagerange{\pageref{firstpage}--\pageref{lastpage}}
\maketitle

\begin{abstract}
Stellar winds and supernova (SN) explosions of massive stars (``stellar feedback'') create bubbles in the interstellar medium (ISM) and insert newly produced heavy elements and kinetic energy into their surroundings, possibly driving turbulence. Most of this energy is thermalized and immediately removed from the ISM by radiative cooling. The rest is available for driving ISM dynamics. In this work we estimate the amount of feedback energy retained as kinetic energy when the bubble walls have decelerated to the sound speed of the ambient medium. We show that the feedback of the most massive star outweighs the feedback from less massive stars. For a giant molecular cloud (GMC) mass of $10^5$~\Msun (as e.g.\ found in the Orion GMCs) and a star formation efficiency of $8$\% the initial mass function predicts a most massive star of approximately $60$~\Msol. For this stellar evolution model we test the dependence of the retained kinetic energy of the cold GMC gas on the inclusion of stellar winds. In our model winds insert $2.34$ times the energy of a SN and create stellar wind bubbles serving as pressure reservoirs. We find that during the pressure driven phases of the bubble evolution radiative losses peak near the contact discontinuity (CD), and thus, the retained energy depends critically on the scales of the mixing processes across the CD. Taking into account the winds of massive stars increases the amount of kinetic energy deposited in the cold ISM from $0.1$\% to a few percent of the feedback energy.
\end{abstract}

\begin{keywords}
ISM: bubbles -- 
hydrodynamics -- 
methods: numerical -- 
stars: massive -- 
stars: winds --
stars: supernovae
\end{keywords}



\section{Introduction}\label{sect:intro}
Massive stars have a dramatic impact on the interstellar medium (ISM): These stars sweep up the ISM with their winds (most prominently in the Wolf-Rayet (WR) phase) and subsequent supernova (SN) explosions and enrich it with freshly-produced heavy elements. This release of momentum and mass -- which will be dubbed ``stellar feedback'' in this paper -- not only affects molecular clouds in star formation regions, but since it is a driver of galactic winds \citep[see e.g.][]{Creasey2013MNRAS429p1922,Nath2013ApJ777p12,Glasow2013MNRAS.434.1151}, also impacts galaxy formation and evolution. Stellar feedback of massive stars is thought to be responsible for the low star formation efficiency in galaxies with molecular gas depletion time-scales of around $10^9$ years \citep{Genzel2015ApJ800p20}. Consequently, a realistic implementation of stellar feedback in numerical simulations is essential to study the dynamics and chemodynamics of the ISM, from the scales of individual molecular clouds to galaxy formation \citep[e.g.][]{Ascasibar2015mnras448p2126,DallaVecchia2012MNRAS426p140,Hopkins2014MNRAS445p581,Kobayashi2001ApJ729p16,Scannapieco2012MNRAS423p1726}.

 In this context, it is important to quantify how much energy massive stars can convert to kinetic energy of the surrounding ISM. Such estimates are relevant to assess in which processes the feedback of massive stars can play a role. For example, such stars are believed to contribute to driving turbulence in the ISM \citep[see e.g.][]{2004RvMP...76..125M} and fuelling galactic winds \citep[see also][]{Hopkins2012MNRAS4213522}. In this work, we check the ratio of retained vs.\ cumulative feedback energy with numerical simulations. The aforementioned ratio will be called ``feedback energy efficiency''. Studies of starburst wind galaxies typically require an overall efficiency of several $10$ per cent (\citealt{Veilleux2005ARA&A43p769,Strickland2009ApJ697p2030}, but compare also \citealt{Glasow2013MNRAS.434.1151}) to power the outflows, while direct observations of super star clusters in M82 lead \citet{Silich2009ApJ700p931} to assume an upper limit of $10$ per cent for this efficiency. 

In the past decades the injection of mass and momentum of (single) SNe into their surroundings has been studied numerically by many authors \citep[e.g.][]{Thornton1998ApJ500p95,Iffrig2015AA576p95,Walch2014arXiv1410.0011,Gatto01052015}. Such studies find that about $3$ to $10$ per cent of the $10^{51}$~erg of the SN energy are retained as kinetic energy of the ambient medium, with a slight dependence on the ambient density and the physics modelled and a strong dependence on the time at which the retained energy is estimated. However, many such studies \citep[e.g.][]{Thornton1998ApJ500p95} ignore the existence of stellar winds and the cavities they create around the star they originate from. Such winds have however been established to profoundly transform the circumstellar medium \citep[compare e.g.][]{Georgy2013AA559A69,vanMarle2012AA547p3,Krause2013AA55049}. For a giant molecular cloud clump with a radius of $4$~pc the 3D study of \citet{Rogers2013} winds and SNe lead to strong leakage of the feedback from this clump. In contrast to this, our 1D study can not cover this leakage and thus applies rather to GMCs like the ones in Orion, which are at least a factor $10$ more massive than this clump.

The impact of the wind of the SN's progenitor star on the ambient medium is clearly observed in many cases: For example by the shell of the progenitor star around SN 1987A reported by \citet{Wampler1990ApJ362p13}, the wind shell of a $25$~\Msun star seen in the SN remnant G296.1--0.5 \citep{Castro2011ApJ734p86} or the stellar-wind envelope seen in SN 2006aj \citep{Sonbas2008AstBu63p228}. 

In the processes studied in this work (we follow the energy injections of winds and supernovae) the bubble produced by the massive star is filled with a hot, dilute gas that contains elements created via stellar nucleosynthesis. It is surrounded by walls of cold, dense swept-up ambient medium. To estimate the spread of the ejecta, it is important to take into account, how well these two media mix. In addition, these mixing processes also affect the cooling physics and consequently impact the feedback energy efficiency. Unfortunately, due to the large range of scales a hydrodynamical treatment of these mixing processes is beyond reach in most simulations. Therefore many chemical evolution models assume an immediate mixing of the SN ejecta in the walls of super-bubbles. However, it is unclear if this is realistic. As pointed out by e.g.\ \citet{TenorioTagle1996AJ111p1641} stellar winds and supernova explosions lead to a two shock structure with a contact discontinuity (CD) separating the well mixed hot material inside the bubble from the swept-up, compressed, heated, radiatively cooling (and thus cold) ambient medium. From two-dimensional numerical simulations of SN ejecta colliding with the swept-up wind material in wind-blown bubbles \citet{TenorioTagle1996AJ111p1641} reports R-T instabilities followed by K-H instabilities due to this collision whereas \citet{Pan2012ApJ756p102} report a stable CD for isotropic ejecta. However, \citet{Pan2012ApJ756p102} note that the omnipresent turbulence in the ISM will lead to instabilities, which in turn enhance the mixing across the CD by increasing the CD surface. Mixing of shell material into the hot bubble gas has been suggested to enhance its density, such that supernova shock waves may then produce the observed temperatures and luminosities when running into such mixing regions \citep{Krause2014apjl794L21,Krause2014AA566A94K}. Unfortunately, the efficiency of mixing across the CD still remains an open question, and presently the mechanism of mixing via droplets produced in the SN receives most attention \citep{Stasinska2007AA471p193,Gounelle2009ApJ694p1,Gounelle2012AA545p4,Boss2012ApJ756p9,Pan2012ApJ756p102}. Examples of micro- and macroscopic processes capable of degrading the CD are briefly discussed in the Appendix in Sect.~\ref{sect:conduction}, \ref{sect:diffusion} and \ref{sect:turbdiff}. In our 1D pilot study the evaporation of cold clumps deep inside a cavity and the leakage of feedback from a structured GMC can not be taken into account. To some extent one can parametrize these effects in terms of an assumed mixing length, if they appear close to the CD. The former creates a larger amount of intermediate density gas, which is directly parametrized by the mixing length. The latter is effectively an enhanced dissipation of energy, which follows from the mixing length by the resulting change in radiative dissipation.

In our work we will thus study the dependence of the feedback energy efficiency on the assumed mixing efficiency. We will also show that pre-existing bubbles at the time of the SN explosion greatly enhance the feedback energy efficiency. Generalising this result indicates that also in superbubbles, where the most massive stars explode first, subsequent SNe become more efficient, since they can take advantage of the low density bubbles produced by the more massive stars.

Our work extends the published stellar feedback energy efficiency models in two aspects: (1) The energy content of the simulations is monitored until the shell is decelerated to the sound speed of the ambient medium (the motivation for this is discussed in Sect.~\ref{sect:simulationsetup}). This is longer than in the work of \citet{TenorioTagle1990MNRAS244p563,TenorioTagle1991MNRAS251p318,TenorioTagle1996AJ111p1641}. (2) Variations of the wind strength with time are taken into account in the feedback model (Sect.~\ref{sect:feedbackmodel}).

We do not include the effects of \textsc{H\,ii} regions or radiation pressure in our models. An estimate of the relative importance of these not included processes in comparison to winds and SNe for different types of stars can be found in the review of \citet{Dale2015NewAR68p1}. Simulations with momentum driven winds \citep[e.g.][]{Ngoumou2015ApJ798p32,Dale2015NewAR68p1} report a dominance of radiation over the wind momentum for similar stellar parameters as our study uses. However, in our models wind bubbles are pressure driven before the SN and wind momentum is also a second order effect. \citet{Freyer2003ApJ594p888}, who also include the pressure increase caused by stellar winds, find that the ionization energy dominates during the first $2$~Myr of the evolution of a $60$~\Msun star in a homogeneous $n_0=20$~\mcmb and $T_0=200$~K medium, but gets comparable to the kinetic energy and thermal energy of hot gas later on. In our models, the cavity-size at the time of the SN explosion influences the retained energy. Including these processes will thus increase the bubble size, which in turn increases the retained energy. However it will not change the finding, that pre-existing bubbles at the time of the SN explosion are an important feature and that the assumed scale length of mixing processes has a strong influence on the retained energy.

This paper will first discuss the implementation of our model (Sect.~2), then proceed to models with SNe only (no stellar winds, Sect.~3). In Sect.~4 we will discuss models with stellar winds and SNe and finally, in Sect.~5 and 6, we will summarize our findings.
\section{Method}
\begin{table}
\begin{tabular}{crrrr}
$\rho$ [g \mcm] & \multicolumn{1}{c}{$t$ [kyr]}  & \multicolumn{1}{c}{$t/t_0$} & \multicolumn{1}{c}{$E_{\rm kin}$ [$10^{50}$~erg]} & \multicolumn{1}{c}{$r$ [pc]} \\ 
\hline \\[-1.0em]
\multicolumn{5}{l}{\citet{Thornton1998ApJ500p95} shell only}\\
\multicolumn{5}{l}{ $T=1\,000$~K, $\Delta x = 0.056$~pc, 3~\Msun}\\
$2.2 \times 10^{-25}$ & 122   & 1\hspace{0.5em}   & $2.14$\hspace{1em}  &  55.8  \\
                     & 1\,590 & 13\hspace{0.5em} & $0.77$\hspace{1em}  & 114.3  \\
$2.2 \times 10^{-24}$ & 34.4  & 1\hspace{0.5em}   & $2.17$\hspace{1em}  &  21.4  \\
                     & 447   & 13\hspace{0.5em}  & $0.75$\hspace{1em}  &  43.0  \\
$2.2 \times 10^{-23}$ & 9.73  & 1\hspace{0.5em}   & $2.33$\hspace{1em}  &   8.2  \\
                     & 126   & 13\hspace{0.5em}  & $0.84$\hspace{1em}  &  16.4  \\
$2.2 \times 10^{-22}$ & 3.06  & 1\hspace{0.5em}   & $2.35$\hspace{1em}  &   3.3  \\
                     & 39.8 & 13\hspace{0.5em}   & $0.76$\hspace{1em}  &   6.6  \\
\hline \\[-1.0em]
\multicolumn{5}{l}{\citet{Thornton1998ApJ500p95} whole SNR}\\
\multicolumn{5}{l}{ $T=1\,000$~K, $\Delta x = 0.056$~pc, 3~\Msun}\\ 
$2.2 \times 10^{-25}$ & 122   & 1\hspace{0.5em}   & $2.73$\hspace{1em}  &  55.8  \\
                     & 1\,590 & 13\hspace{0.5em} & $0.78$\hspace{1em}  & 114.3  \\
$2.2 \times 10^{-24}$ & 34.4  & 1\hspace{0.5em}   & $2.74$\hspace{1em}  &  21.4  \\
                     & 447   & 13\hspace{0.5em}  & $0.84$\hspace{1em}  &  43.0  \\
$2.2 \times 10^{-23}$ & 9.73  & 1\hspace{0.5em}   & $2.67$\hspace{1em}  &   8.2  \\
                     & 126   & 13\hspace{0.5em}  & $0.76$\hspace{1em}  &  16.4  \\
$2.2 \times 10^{-22}$ & 3.06  & 1\hspace{0.5em}   & $2.61$\hspace{1em}  &   3.3  \\
                     & 39.8 & 13\hspace{0.5em}   & $0.80$\hspace{1em}  &   6.6  \\
\hline \\[-1.0em]
\multicolumn{5}{l}{$T=1\,000$~K, $\Delta x = 0.004$~pc, $r_{\rm f} = 1.5$~pc, 3~\Msun}\\
$2.2 \times 10^{-25}$ & 96.5   & 1\hspace{0.5em}       & $2.84$\hspace{1em}  & 47.5 \\
                     & 1\,245.5 & 13\hspace{0.5em}    & $0.82$\hspace{1em}  & 106.2 \\
$2.2 \times 10^{-24}$ &  28.0 & 1\hspace{0.5em}        & $2.77$\hspace{1em}  & 18.6 \\
                     & 364.0 & 13\hspace{0.5em}       & $0.78$\hspace{1em}  & 39.4 \\
$2.2 \times 10^{-23}$ &   8.0 & 1\hspace{0.5em}        & $2.69$\hspace{1em}  &  7.3 \\
                     & 104.0 & 13\hspace{0.5em}       & $0.74$\hspace{1em}  & 15.1 \\
$2.2 \times 10^{-22}$ &   2.5 & 1\hspace{0.5em}        & $3.23$\hspace{1em}  &  3.1 \\
                     &  32.5 & 13\hspace{0.5em}       & $0.66$\hspace{1em}  &  6.0 \\
\hline \\[-1.0em]
\multicolumn{5}{l}{$T=1\,000$~K, $\Delta x = 0.004$~pc, $r_{\rm f} = 0.3$~pc, 11~\Msun}\\
$2.2 \times 10^{-25}$ & 100.5   & 1\hspace{0.5em}       & $2.68$\hspace{1em}  & 49.4 \\
                     & 1\,306.5 & 13\hspace{0.5em}    & $0.80$\hspace{1em}  & 103.5 \\
$2.2 \times 10^{-24}$ &  30.0 & 1\hspace{0.5em}        & $2.68$\hspace{1em}  & 19.1 \\
                     & 390.0 & 13\hspace{0.5em}       & $0.73$\hspace{1em}  & 38.3 \\
$2.2 \times 10^{-23}$ &   9.0 & 1\hspace{0.5em}        & $2.81$\hspace{1em}  &  7.5 \\
                     & 104.0 & \hspace{0.5em}         & $0.72$\hspace{1em}  & 15.0 \\
                     & 117.0 & 13\hspace{0.5em}       & $0.66$\hspace{1em}  & 15.6 \\
$2.2 \times 10^{-22}$ &   3.0 & 1\hspace{0.5em}        & $3.03$\hspace{1em}  &  3.0 \\
                     &  39.0 & 13\hspace{0.5em}       & $0.59$\hspace{1em}  &  6.2 \\
\end{tabular}   
\caption{\small Comparison of the retained kinetic energy ($E_{\rm kin}$) of SNe ($10^{51}$~erg inserted at $t=0$) in homogeneous media. For all models $E_{\rm kin}$ and the bubble radius ($r$) were evaluated at the time of maximal luminosity ($t_0$) and after $13\,t_0$, which is the end of the simulations in \citet{Thornton1998ApJ500p95}. The resolution ($\Delta x$) and the state of the ambient medium ($T$, $\rho$) are varied. Since the bubble pressure at $t_0$ is much higher than the ambient pressure, the efficiency of the $1\,000$~K model is comparable to the $40$~K model. $40$~K is the equilibrium temperature for a density of $2.2 \times 10^{-22}$~g~\mcmb for the applied cooling function. For the ambient medium in the $1\,000$~K model an artificially stable phase had to be implemented in the cooling model. $t_0$ also depends on the size of the feedback region ($r_{\rm f}$) and on the kinetic to thermal energy ratio. Therefore three SN models are shown: the model of \citet{Thornton1998ApJ500p95} with a mass loss of $3$~\Msol, our standard SN prescription and purely thermal energy injection in the $40$~K models.}
\label{tab:thornton}
\end{table}
\begin{table}
\begin{tabular}{crrrr}
$\rho$ [g \mcm] & \multicolumn{1}{c}{$t$ [kyr]}  & \multicolumn{1}{c}{$t/t_0$} & \multicolumn{1}{c}{$E_{\rm kin}$ [$10^{50}$~erg]} & \multicolumn{1}{c}{$r$ [pc]} \\ 
\hline \\[-1.0em]
\multicolumn{5}{l}{$T=40$~K, $\Delta x = 0.032$~pc, 0~\Msun}\\ 
$2.2 \times 10^{-22}$ &   2.5  & 1\hspace{0.5em}        & $2.89$\hspace{1em}  & 2.8    \\
                     &  32.5  & 13\hspace{0.5em}       & $0.61$\hspace{1em}  & 5.9    \\
                     &  39.0  &                        & $0.53$\hspace{1em}  & 6.2    \\
\hline \\[-1.0em]
\multicolumn{5}{l}{$T=40$~K, $\Delta x = 0.016$~pc, 0~\Msun} \\
$2.2 \times 10^{-22}$ &   3.0  & 1\hspace{0.5em}        & $2.95$\hspace{1em}  & 3.0    \\
                     &  32.5  &                        & $0.63$\hspace{1em}  & 5.9    \\
                     &  39.0  & 13\hspace{0.5em}       & $0.55$\hspace{1em}  & 6.2    \\
\hline \\[-1.0em]
\multicolumn{5}{l}{$T=40$~K, $\Delta x = 0.008$~pc, 0~\Msun} \\
$2.2 \times 10^{-22}$ &   3.0  & 1\hspace{0.5em}        & $2.97$\hspace{1em}  & 3.0    \\
                     &  32.5  &                        & $0.66$\hspace{1em}  & 5.9    \\
                     &  39.0  & 13\hspace{0.5em}       & $0.58$\hspace{1em}  & 6.2    \\
\hline \\[-1.0em]
\multicolumn{5}{l}{$T=40$~K, $\Delta x = 0.004$~pc, 0~\Msun} \\
$2.2 \times 10^{-22}$ &   3.0  & 1\hspace{0.5em}        & $2.96$\hspace{1em}  & 3.0    \\
                     &  32.5  &                        & $0.68$\hspace{1em}  & 5.9    \\
                     &  39.0  & 13\hspace{0.5em}       & $0.59$\hspace{1em}  & 6.2    \\ 
\end{tabular}   
\contcaption{}
\label{tab:thornton:2}
\end{table}
\begin{table*}
 \begin{minipage}{\textwidth}
 \centering
\begin{tabular}{ccccccccc} 
 $\Delta x$ & $\Delta x$ & SN & wind & thermal  & $a$ &  $\epsilon$ ($v_{\rm sh}=c_{\rm s}$) &  $\epsilon_{\rm k}$ (wind) &  $\epsilon_{\rm t}$ (wind)\\
\,[pc] & [$10^{16}$~cm] &[$10^{51}$~erg]&[$2.34\times 10^{51}$~erg]& conduction &  & [$10^{51}$~erg] & [$10^{51}$~erg] &  [$10^{51}$~erg] \\
\hline \\[-1.0em]
0.032 & 10.0 & yes & no          & no & 0 &  0.0011 & - & - \\ 
0.016 &\hspace{0.5em}5.0 & yes & no          & no & 0 &  0.0011 & - & - \\ 
0.008 &\hspace{0.5em}2.5 & yes & no          & no & 0 &  0.0011 & - & - \\ 
\hline \\[-1.0em]                             
0.032 & 10.0 & no  & yes         & no & 0 &  0.0213 & 0.0884 & 0.4981 \\ 
0.016 &\hspace{0.5em}5.0 & no  & RW & no & 0 &  0.0231 & 0.0896 & 0.4981 \\ 
\hline \\[-1.0em]
0.064 & 10.0 & yes & yes             & no & 0 &  0.0265 & 0.1027 & 0.5422 \\ 
0.032 & 10.0 & yes & yes             & no & 0 &  0.0271 & 0.0884 & 0.4981 \\ 
0.016 &\hspace{0.5em}5.0 & yes & RW  & no & 0 &  0.0304 & 0.0896 & 0.4981 \\
0.016 &\hspace{0.5em}5.0 & yes & yes & no & 0 &  0.0365 & 0.1136 & 0.6019 \\ 
0.008 &\hspace{0.5em}2.5 & yes & yes & no & 0 &  0.0475 & 0.1340 & 0.6859 \\ 
0.004 &\hspace{0.5em}1.25 & yes & yes & no & 0 &  0.0620 & 0.1598 & 0.7756 \\ 
\hline \\[-1.0em]                          
0.032 &10.0 & yes & yes         & no & 1 &  0.0710 & 0.1841 & 0.8286 \\ 
0.016 &\hspace{0.5em}5.0 & yes & yes         & no & 1 &  0.0791 & 0.1947 & 0.8696 \\ 
0.008 &\hspace{0.5em}2.5 & yes & yes         & no & 1 &  0.0904 & 0.2076 & 0.9113 \\ 
\hline \\[-1.0em]                          
0.032 & 10.0 & yes & yes        & yes & 0 &  0.0244 & 0.0827 & 0.4549 \\ 
0.016 &\hspace{0.5em}5.0 & yes & yes        & yes & 0 &  0.0302 & 0.1014 & 0.5570 \\
\hline \\[-1.0em]                      
0.032 & 10.0 & yes & yes    & extreme & 0 &  0.0094 & 0.0329 & 0.1915 \\ 
0.016 &\hspace{0.5em}5.0 & yes & yes    & extreme & 0 &  0.0098 & 0.0353 & 0.2211 \\ 
\hline \\[-1.0em]
0.032 & 10.0 & yes & CW  & no & 0 &  0.0293 & 0.0932 & 0.2070 \\                      
\hline
\end{tabular}
\caption[Grid of models.]{Grid of models. The ambient medium in all models has a density of $2.2 \times 10^{-22}$~g~\mcmb and a pressure of  $1.48 \times 10^{-12}$~erg~\mcmb corresponding to an equilibrium temperature of approximately $40$~K. $\Delta x$ is the cell size in the simulation. Despite the lower ambient temperatures the three uppermost models without winds are comparable to \citet{Thornton1998ApJ500p95}. The major difference is that we followed these models for a substantially longer time than \citet{Thornton1998ApJ500p95} and thus observe lower efficiencies at the end of our simulations. For models with a supernova explosion (``yes'' in column 3), $10^{51}$~erg and $11$~\Msun of ejecta were inserted after $4.8596$~Myr. For simulations with stellar winds (``yes'' in column 4) the \citet{Geneva2011} model for a rotating $60$~\Msun star and the wind velocities summarized in \citet{Voss2009} were used. In total this stellar wind inserts $2.34 \times 10^{51}$~erg. The constant wind model (``CW'' in column 4) inserts the same total wind energy at a constant rate. To check the influence of the resolution on the feedback energy efficiency of the supernova explosion, simulations with lower resolution were resampled directly before the SN (indicated as ``RW'' in column 4), since the efficiency during the wind phase also depends on the resolution. The slightly higher kinetic energy in the rescaled model at the end of the wind phase is due to smooth interpolation. $\epsilon$ lists the kinetic energy in $10^{51}$~erg when the cell with the highest density is decelerated to the sound speed of the ambient medium. $\epsilon_{\rm k}$ and $\epsilon_{\rm t}$ list the retained kinetic and thermal energy at the end of the wind phase (in units of $10^{51}$~erg). ``Extreme'' thermal conduction mimics a very efficient diffusion process by increasing $\kappa$ by $14$ orders of magnitude. The parameter $a$ describes a density threshold below which radiative cooling is no longer taken into account. This decreases the energy losses due to mixing of gas across the CD. The threshold density $a$ is normalized to the density of the ambient medium. The table shows that {\bf higher} efficiencies are reached for higher resolutions, thus the higher maximal densities are outweighed by the smaller amount of mixing across the CD in the higher resolved simulations. Whereas in lower resolved simulations a decrease of the efficiency with increasing resolution is found, since the cell near the CD is too large to reach high enough densities or temperatures due to the mixing across the CD to suffer substantial energy radiative losses at every time-step.}
\label{tab:newGrid}
\end{minipage}
\end{table*}
In our hydrodynamic simulations a massive star (Sect.~\ref{sect:feedbackmodel}) is placed in a homogeneous medium (Sect.~\ref{sect:cloud}) where it first produces a stellar wind bubble and subsequently undergoes a SN explosion. 

The deployed numerical methods are suitable to treat a contact discontinuity (CD) separating two distinct phases of the ISM inside the bubble: a hot dilute\footnote{several orders of magnitude below the ambient density, $10^6$~K or hotter} wind phase, which cannot cool due to its low density and a cold, denser\footnote{more than a factor $4$ denser than the ambient medium, $10$~K} phase, which also does not cool strongly, because it is too close to the cooling-heating equilibrium temperature to cool efficiently. An analogous behaviour (at higher temperatures) can be e.g.\ seen in the cooling curves presented by \citet{Sutherland1993ApJS88p253}, which show a strong decrease of $\Lambda(n,T)$ below $10\,000$~K. 

Due to the presence of the CD, the feedback energy efficiency depends on the mixing of these two gas phases and is thus influenced by the spatial resolution and the diffusivity of the numerical scheme. This unavoidable diffusivity can be related to physical diffusive processes, which are discussed in Appendix~\ref{CDdiffusion}. 

We study the stellar feedback energy efficiency in 1D to be able to conduct a sensitivity study covering a wide range of parameters at high resolution.
\subsection{Setup of the simulations}\label{sect:simulationsetup}
The 1D spherically symmetric simulations were carried out with the Eulerian mesh code {\sc Pluto} \citep{Mignone2007ApJS170p228}. Our modifications of the code are a cooling-heating prescription as described in \citet{Ntormousi2011ApJ}, which allows for a multi-phase ISM and is based on the cooling-heating function of \citet{Wolfire1995ApJ443p152},  a time dependent feedback of a $60$~\Msun star (Sect.~\ref{sect:feedbackmodel}), a minimal density to numerically stabilize the very dilute hot zones inside the bubbles, and a threshold density below which radiative cooling is not taken into account. The latter can be used to stabilize cells near the CD (Appendix, Sect.~\ref{sect:separate}). Most of our models used an approximate Riemann solver that is able to represent contact surfaces particularly well (HLLC Riemann solver, see also Appendix Sect.~\ref{sect:solver}), Linear TVD and second order Runge-Kutta time-stepping. For comparison we show the results of simulations with different solvers in the Appendix Sect.~\ref{sect:solver}. If nothing else is mentioned, the ISM in our models can cool down to $10$~K if the density is above $0.01$~\pccm. 

The spatial resolution of the simulations in our set of models is up to $250$ cells per parsec (The highest meaningful number of cells per parsec is discussed in Appendix Sect.~\ref{sect:mfp}.). We started with a computational box with an edge length of $5$~pc and monitored during the simulation if at least $100$ undisturbed cells of ambient medium were left. If the number of undisturbed cell became too small we added another $5$~pc of undisturbed medium to the computational box.

Our simulations stop when the swept-up shell is decelerated to the sound speed of the ambient medium. We argue that at this point turbulent motions will lead to break-up of the shell and very efficient mixing (and energy deposition) in the ambient medium.

\subsection{Cloud material}\label{sect:cloud}
The standard assumptions for the cloud material in this study are solar metallicity and a density of $2.2 \times 10^{-22}$~g~\mcm. With our chosen cooling-heating prescription the cooling-heating equilibrium for this density is reached at a pressure of  $1.48 \times 10^{-12}$~erg~\mcmb  corresponding to an equilibrium temperature of approximately $40$~K. As shown later (Table~\ref{tab:thornton}), the exact temperature does not play an important role. The number density ($n$) of $\sim 100$~\mcmb corresponds to the average density of molecular cloud complexes \citep{Murray2011ApJ729p133}. It is known that molecular clouds exhibit a fractal structure. These inhomogeneities and the high density clumps and cores will be addressed in future work. 
 
 The median value of the number density of H$_2$ in the galactic ring survey \citep{Roman2010ApJ723p492} is $231$~\mcm. However, this survey is likely biased towards high density regions, since it is based on $4\sigma$ $^{13}$CO contours. Similar techniques led to a factor of $10$ lower densities in \citet{Heyer2009ApJ699p1092}. Thus, our assumption of $\sim 100$~\mcmb lies well in the plausible region of average densities in molecular clouds.
\subsection{Stellar feedback}\label{sect:feedbackmodel}
\begin{figure}
\includegraphics[width=\columnwidth, trim = 2mm  0mm 1mm 0mm, clip]{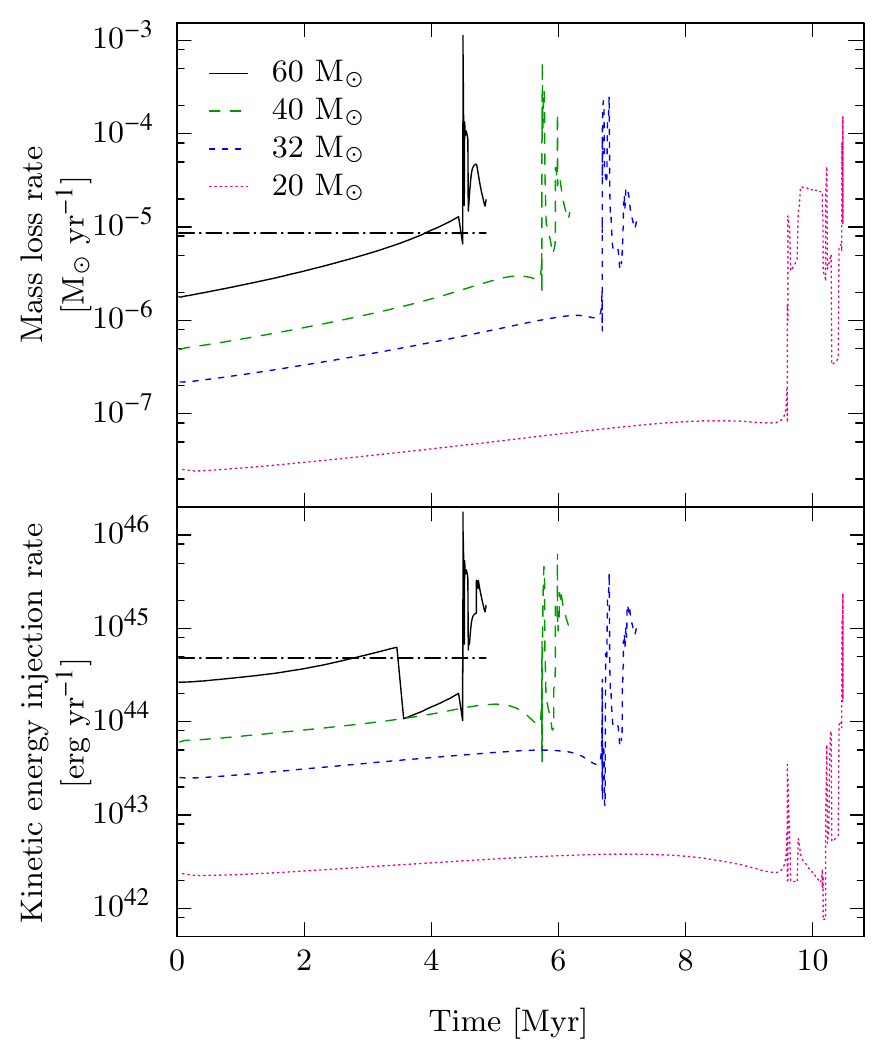}
\caption[Stellar feedback of massive stars]{Stellar feedback of massive stars. The kinetic energy output of massive stars is shown in the lower panel according to the tabulated mass loss rates of the rotating stellar models of \citet{Geneva2011} (upper panel) combined with the estimated terminal wind velocities as summarized in \citet{Voss2009}. The dot-dashed line indicates a constant wind with the same net energy and mass input as the $60$~\Msun star model with time dependent power input. Before the SN (at $t=4.86$~Myr) the mass loss rate is $8.65\times 10^{-6}\,{\rm M_{\odot}\,yr^{-1}}$ and the energy injection rate is $4.8\times 10^{44}$~erg~yr$^{-1}$. In total the $60$~\Msun star injects $2.34\times 10^{51}$~erg into its surroundings during the wind phase.} 
\label{fig:simplemodel}
\end{figure}
The stellar feedback is calculated from the mass loss rate and the surface abundances of the rotating models of \citet{Geneva2011}. The surface abundances are used to determine the WR type and for each type the wind velocity as summarized in \citet{Voss2009} was applied. Fig.~\ref{fig:simplemodel} shows the strength of the stellar feedback with time for different types of massive stars.  It illustrates that the most massive star remaining in a population dominates the feedback energy \citep[see also e.g.][]{Oey2005AAS207p6001}.

 A plausible mass of this star can be estimated from the molecular clouds in the Milky Way: The assumption that about $8$ per cent \citep{Murray2011ApJ729p133} of the molecular cloud mass are converted to stars leads to a cluster mass of $8\times 10^3$~\Msun for a molecular cloud of $10^5$~\Msol. In the galactic ring survey \citep{Roman2010ApJ723p492} $\sim 18$ per cent and in the list of \citet{Heyer2009ApJ699p1092} $\sim 31$ per cent of the galactic molecular clouds are estimated to be more massive than $10^5$~\Msol. \citet{Weidner2013MNRAS434p84} find a most massive star of $\sim 60$~\Msun for a cluster mass of $8\times 10^3$~\Msun with their polynomial fit to the observed most massive stars as a function of the cluster mass.  Since a good fraction of the GMCs can harbour most massive stars of $60$~\Msol, we focus on the feedback energy efficiency of a star of this mass.  The stellar winds in this model play an important role, since they insert $2.34$ times the SN energy into the ambient ISM (Fig.~\ref{fig:simplemodel}). This wind-to-SN ratio is larger than in \citet{Voss2009}, since we consider individual massive stars whereas \citet{Voss2009} are interested in OB-associations. In groups of stars, less massive stars lower the ratio of wind energy to SN energy if a canonical SN energy of $10^{51}$~erg is assumed.

In the SN blast of the $60$~\Msun star, $11$~\Msun of material are ejected (see Sect.~\ref{sect:FB:model}).  In our model these ejecta are initially homogeneously distributed over a small sphere of radius $r_{\rm f}=0.32$~pc, which we will refer to as the ``feedback region''. Since the spherically symmetric grid in all our simulations starts at $0.032$~pc, the model with the lowest resolution presented here has $9$ grid zones inside the feedback region. Test simulations showed that the size of this feedback region does not influence the results as long as it is small enough to be fully contained in the wind bubble. All models including winds respect this condition.

However, at the absence of a wind, the size of the feedback region can influence the kinetic to thermal energy ratio after $13$ times of maximal luminosity ($t_0$). \citep[$13 t_0$ are called $t_{\rm f}$ in][and these simulations end at $t_{\rm f}$.]{Thornton1998ApJ500p95} The time of maximal luminosity is defined as the moment when the largest energy losses due to radiative cooling occur in the simulation. (Despite its name, $t_0$ should not be confused with the time of the maximum in the SN light curve, which is caused by radioactive decays). In these cases we performed convergence studies to decide the optimum size of the feedback region (Sect.~\ref{sect:FB:model}).

The time dependent stellar mass loss of the wind is inserted homogeneously in the feedback region. The time integrated mass loss is found by trapezoidal integration in the tabulated data of \citet{Geneva2011}. A part of the wind energy is inserted as kinetic energy in the feedback region by adding gas mass to cells with non-zero gas-velocity. The rest of the wind energy is added as thermal energy. Models that inserted all feedback energy as kinetic energy by imposing a linear velocity profile in the feedback region (to mimic the velocity of a Sedov-Taylor sub-grid model) or a constant velocity (to mimic a free streaming wind region) led to the same results regarding the feedback energy efficiency. 

However, adding kinetic energy produced artefacts in the wind along the grid in test simulations with $2$ or $3$ dimensions. Since the models presented in this study are only a subset of a larger set of models, including models with $2$ or $3$ dimensions, we prefer to insert the major part of the feedback energy thermally. In models with SNe and without winds the thermal energy fraction in the SN was either $72$ per cent \citep[obtained from the self-similar solution of the Sedov-Taylor problem, see e.g][]{Chevalier1974ApJ188p501} if a linear velocity profile was imposed on this region or $100$ per cent thermal energy otherwise.
\section{Results: SNe without progenitor winds} \label{sect:SNonly} 
The supernova models discussed in this section do not take the stellar winds of the progenitor star into account. Hence at the time of the SN explosion the ambient ISM in these models is homogeneous without pre-existing stellar wind bubbles. These models do not only provide a consistency check of our setup with published feedback energy efficiencies \citep{Thornton1998ApJ500p95,TenorioTagle1990MNRAS244p563}, but also go beyond them, since our simulations follow the SN shell until it has been decelerated to the sound speed of the ambient medium, which is substantially longer than the simulations in the aforementioned works were monitored. This allows us to study the full evolution of the energy deposition in the surrounding ISM.

 Details on the evolution of the SN bubble can be found in Appendix~\ref{sect:SNevolution}.
\subsection{Grid of models}
We ran a large number of simulations varying the ambient densities from $2.2 \times 10^{-25}$~g~\mcmb to $2.2 \times 10^{-22}$~g~\mcm. To compare to previous work, we keep the initial temperature at $1\,000$~K (see Table~\ref{tab:thornton}) in order to check the influence of the ambient pressure and also of the spatial resolution on the results. A subset of these models (no stellar wind, ambient density $2.2 \times 10^{-22}$~g~\mcm) is also shown in the uppermost part of Table~\ref{tab:newGrid}. In these models we find very low feedback efficiencies, but models converge nicely with resolution.

Due to the sharp discontinuity between the hot bubble and the cold shell, low order interpolation functions and the exact Riemann solver (``two shock'' in {\sc Pluto}, see Appendix Sect.~\ref{sect:solver}) have to be used to avoid numerical effects near the contact discontinuity, which in turn would cause negative pressures and spurious energy gains.

\begin{figure}
 \includegraphics[width=\columnwidth, trim = 2mm  1mm 1mm 1mm, clip]{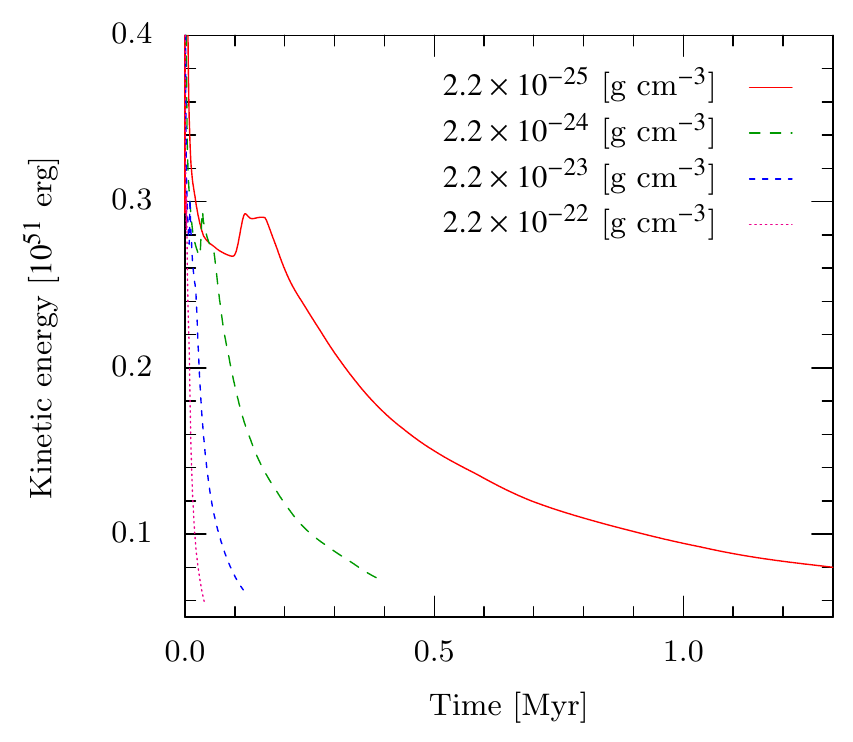}\\
 \includegraphics[width=\columnwidth, trim = 2mm  1mm 1mm 1mm, clip]{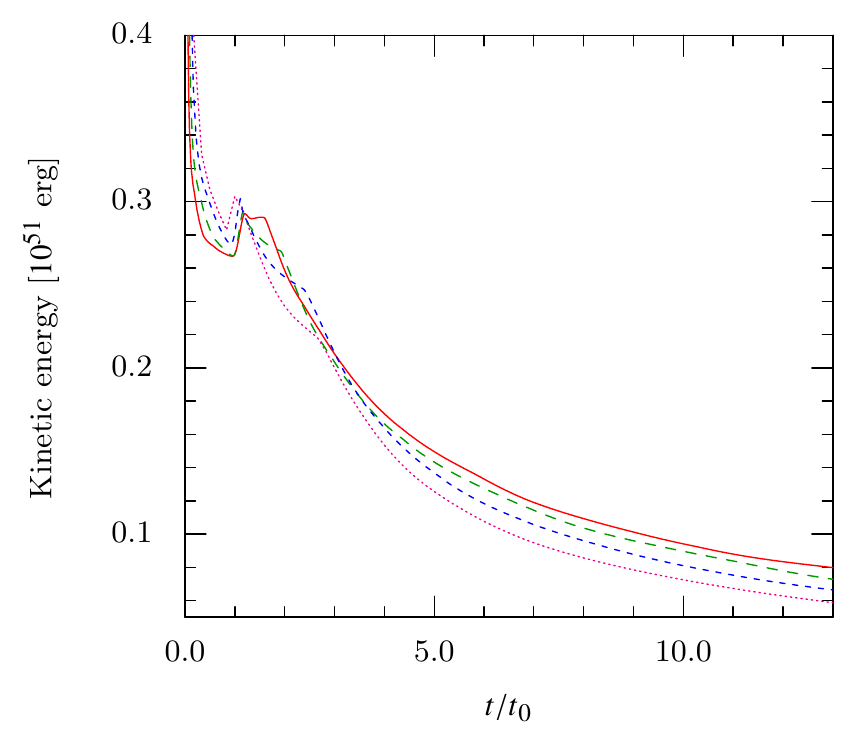}
 \caption{Retained kinetic energy of a SN in a homogeneous medium with a temperature of $1\,000$~K in units of SN energies ($E_{\rm SN}=10^{51}$~erg). For this simulation an artificially stable ISM phase at the temperature and the density of the ambient medium was used. In our simulations a lower feedback energy efficiency in denser media is observed. This figure shows the numerical simulations for a SN with a thermal energy fraction of $0.7 E_{\rm SN}$, a mass loss of $11$~\Msun and a feedback region radius of $0.3$~pc. Both panels show the same models for different ambient densities: The time axis in the lower panel is scaled with the time of maximal luminosity, $t_0$. }
 \label{fig:Thornton:Ek+t}
\end{figure}
\begin{figure*}
 \includegraphics[width=\textwidth, trim = 2mm  1mm 1mm 1mm, clip]{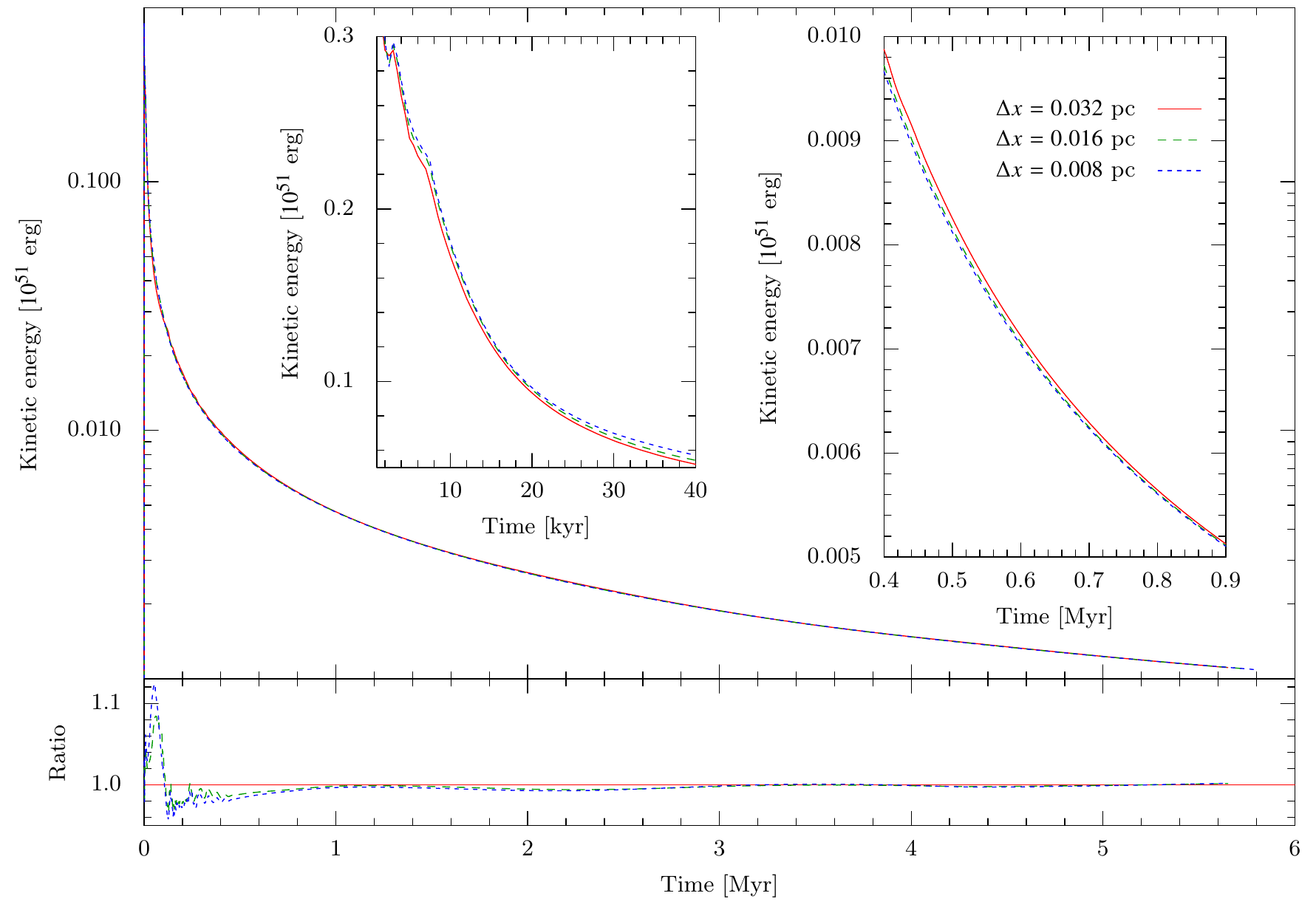}
\caption{Retained kinetic energy (in units of canonical SN energies ($10^{51}$~erg)) of a SN in a homogeneous medium ($T=40$~K and $\rho_0=2.2 \times 10^{-22}$~g~\mcm) with purely thermal energy input. The energy is quickly lost via radiative cooling, but the shell needs more than $5.6$~Myr to decelerate to the sound speed of the ambient medium. The lower panel compares the retained kinetic energy to the retained kinetic energy in the lowest resolution model. After a Myr the results for different resolutions are very well converged. In the kinetic energy ratios it can be seen that higher resolution models lose less energy in the pressure driven phase due to the smaller cooling region at the sides of the shell [in this phase the dashed lines are above the solid line in the lower panel], but make up in the momentum conserving phase [dashed line below solid line]. The lines end when the shell is decelerated to the sound speed of the ambient medium. The left insert shows the pressure driven phase. The convergence of the retained energies at different resolutions can be seen in the right insert and in the lower panel depicting the energy content of the models divided by the energy content of the model with the lowest resolution at the same time.}
\label{fig:NOwind:Ek+t}
\end{figure*}
\begin{figure}
\includegraphics[width=\columnwidth, trim = 2mm  1mm 1mm 1mm, clip]{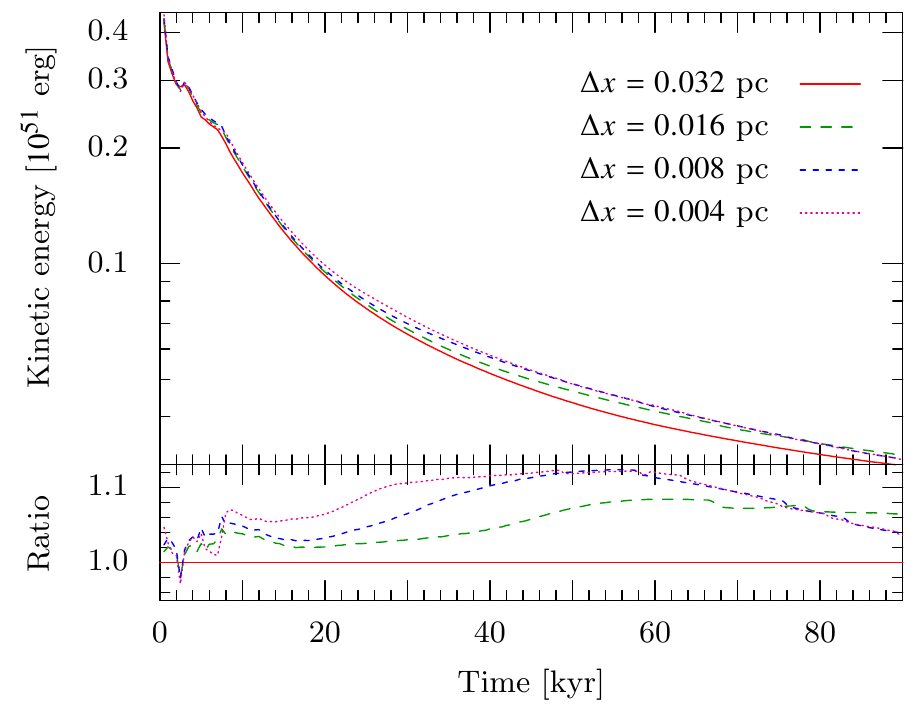} 
\caption{Zoom in of Fig.~\ref{fig:NOwind:Ek+t}. In this plot the highest resolution model is added, which was stopped after $20 t_0$.}
\label{fig:NOwind:Ek+t:c}
\end{figure}
\subsection{Comparison to previous work}\label{sect:thornton}
A very well studied case of a SN explosion in the literature \citep{Thornton1998ApJ500p95,TenorioTagle1990MNRAS244p563} is the deposition of $E_{\rm SN}=10^{51}$~erg into a homogeneous ambient medium with a number density of $n_0=1$~\mcm.

The study of \citet{Thornton1998ApJ500p95} covers also ambient densities better matching to GMCs than the aforementioned  $n_0=1$~\mcmb medium, which models the warm phase of the ISM. To compare models with different ambient densities, they normalized the simulation times with the corresponding ``times of maximal luminosity'' ($t_0$, defined in Sect.~\ref{sect:feedbackmodel}).

 \citet{Thornton1998ApJ500p95} found a feedback energy efficiency of $\sim 10$ per cent after $13\, t_0$ for a wide range of ISM number densities ($n = 0.001$ to $1\,000$~\mcm) and metallicities ($\log \left(Z/Z_{\odot}\right) = -3.0$ to $0$). In their model for a SN explosion without prior stellar wind bubble in a homogeneous ambient medium with $\rho_0=2.2\times 10^{-22}$~g~\mcm, solar metallicity and a temperature of $1\,000$~K \citet{Thornton1998ApJ500p95} find a feedback energy efficiency of about $8$ per cent after $13$ times of maximal luminosity ($t_{0}$). At this time we find similar feedback energy efficiencies (Table~\ref{tab:thornton}). However, our models show a slightly stronger density dependence of the feedback energy efficiency: Fig.~\ref{fig:Thornton:Ek+t} plots the evolution of the retained kinetic energy as a function of time in Myr in the top panel and normalized to $t_{0}$, which is larger for lower ambient densities, in the lower panel. Table~\ref{tab:thornton} and Fig.~\ref{fig:Thornton:Ek+t} also contain simulations with lower ambient densities than our standard model to simplify the comparison to \citet{Thornton1998ApJ500p95}.

In contrast to \citet{Thornton1998ApJ500p95} who stop the simulations after $13\,t_0$, which is in most models shortly after the transition to the momentum conserving phase, we monitor the simulations until the shell velocity has decreased to the sound speed of the ambient medium. We assume that the remaining kinetic energy will then be dissipated by the ambient medium. Fig.~\ref{fig:NOwind:Ek+t} and Table~\ref{tab:newGrid} show that the model with an ambient density of $2.2 \times 10^{-22}$~g~\mcmb retains just $0.11$ per cent of the SN feedback energy at this time. This efficiency is much smaller than usually assumed for SN feedback.

\subsection{Impact of the ambient pressure}
Since the cooling-heating equilibrium in our chosen cooling prescription predicts an equilibrium temperature of $40$~K for a density of $2.2 \times 10^{-22}$~g~\mcmb also models with this temperature of the ambient medium were added to Table~\ref{tab:thornton}. Comparing these models to the $T=1\,000$~K models shows that the ambient pressure has only a minor effect. The changes in bubble size and kinetic energy are less than one per cent and would thus be invisible in Table~\ref{tab:thornton}. As expected, a higher ambient pressure leads to a slightly smaller bubble. However, this is a very small effect. Overall the resolution  and the implementation of the SN are more important, as can be seen from the models with cell sizes of $0.004$~pc in Table~\ref{tab:thornton}.

\subsection{Impact of the feedback model}\label{sect:FB:model}
The supernova implementation of \citet{Thornton1998ApJ500p95} assumes a mass loss of $3$~\Msun and an energy input $E_{\rm SN}$ of $10^{51}$~erg. They insert $6.9$ per cent of the SN energy via thermal energy and the rest via a linear velocity profile in a region of $1.5$~pc radius. In the rotating $60$~\Msun star model of \citet{Geneva2011} the stellar mass at the point of the SN explosion is $18$~\Msun (the rest of the mass was lost via winds). Assuming a generic remnant mass of $7$~\Msun \citep[like e.g.][]{Voss2009} leads to the ejection of $11$~\Msun of material in the SN blast. 

In our study the radius ($r_{\rm f}$) of the feedback region was reduced until the choice of the kinetic to thermal energy ratio in the SN blast changed the retained kinetic energy ($\epsilon_{\rm k}$) at $t_{\rm f}=13\, t_0$ in the model with the highest ambient density by less than one per cent (of $\epsilon_{\rm k}\left(t_{\rm f}\right)$) in the model with the highest ambient density (Table~\ref{tab:thornton}). Since the bubble size of a Sedov blast is proportional to $\rho^{-1/5}$, models with higher ambient medium density are more sensitive to the too large feedback region problem. In our study this happened at $r_{\rm f}=0.32$~pc. Increasing the feedback region radius to $1.5$~pc decreases the kinetic energy by $\sim 3$ per cent and increases the bubble size by $\sim 0.5$ per cent at $13\, t_0$.
 
 The thermal energy fraction of the SN energy in our $1\,000$~K model is $72$ per cent (Sect.~\ref{sect:feedbackmodel}). In the $40$~K model shown in Table~\ref{tab:thornton} all SN energy was inserted via thermal energy, which leads to a slightly different kinetic to thermal energy ratio in the early phase than models in which the energy fractions at the SN blast are chosen according to the Sedov-Taylor solution.
\subsection{Convergence}
Fig.~\ref{fig:NOwind:Ek+t} and \ref{fig:NOwind:Ek+t:c} show that the feedback energy efficiency of the $T_{\rm eq}=40$~K models without wind is converged for all resolutions ($0.004$ to $0.032$~pc). The retained kinetic energy converges as soon as the shell has cooled to the equilibrium temperature and the dominant radiative cooling losses occur in the spatially well resolved newly swept-up compressed and heated gas at the outside of the shell. At this time the pressure in the swept-up shell is already larger than the pressure inside the bubble and the gas is heated to its equilibrium value due to the expansion of the gas at the inner side of the shell. All resolutions show a feedback kinetic energy efficiency of $0.11$ per cent when the shell speed reaches the sound speed of the ambient medium.
\section{Results: SN blast in a cavity}
\begin{figure}
\includegraphics[width=\columnwidth, trim = 2mm  0mm 1mm 1mm, clip]{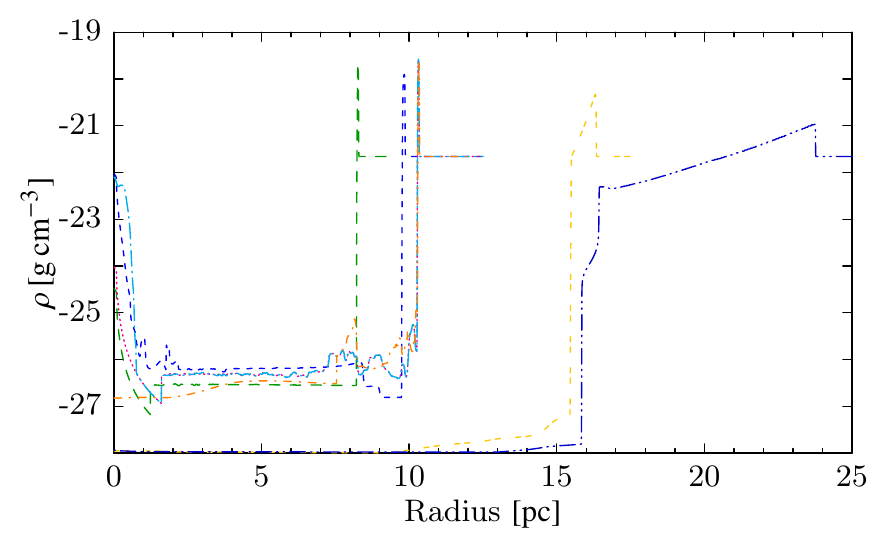}
\caption{Density structure at different stages of the evolution of the model shown in Fig.~\ref{eq+t+nh100x200}. The colours and line styles are the same as in the aforementioned figure. 
Long dashes: start of the WR phase. 
Short dashes: maximal mass loss due to the wind.
Dotted: end of the wind phase.
Long dash dotted: maximal mass loss due to the SN. 
Short dash dotted: $t_0$.
Double dashes: start of the momentum driven phase.
Long dash, double dots: $v_{\rm sh}=c_{\rm s}$.
}
\label{nh100x200+rho}
\end{figure}
Since the progenitor stars of SNe have strong stellar winds, SN explosions always happen inside wind-blown bubbles. In this section we show that this is not a detail, but a very important feature of the model, since it strongly influences the feedback efficiency.

\citet{TenorioTagle1990MNRAS244p563} found feedback efficiencies of $50$ to $70$ per cent for SNe exploding in bubbles blown by a constant WR wind with a mass loss rate of $\dot{M}=3\times 10^{-5}$~\Msol~yr$^{-1}$ and a terminal velocity of $1\,000$~km~s$^{-1}$ into a homogeneous medium with a number density $n=1$~\mcmb and a temperature of $100$~K. In their study the wind phase ends as soon as the wind bubble has reached a predefined diameter. 

The bubbles considered in \citet{TenorioTagle1990MNRAS244p563,TenorioTagle1991MNRAS251p318} and \citet{Rozyczka1993MNRAS261p674} have radii of up to $16$~pc at the time of the SN explosion. These diameters, motivated in these works by observations, are smaller than what we get by applying the more realistic wind models of \citet{Voss2009} to the rotating stellar models of \citet{Geneva2011}. For instance, in a medium $100$ times denser ($n=100$~\mcmb and $T=100$~K), even the least massive star able to produce a SN creates a $13.6$~pc bubble. This probably means that the feedback acts on higher mean densities than those considered in the past, so we put our emphasis on $n=100$~\mcmb models.

The ambient density plays an important role for the feedback energy efficiency: Models with higher ambient densities have lower feedback energy efficiencies (Fig.~\ref{fig:Thornton:Ek+t}, Table~\ref{tab:thornton}). Table~\ref{tab:newGrid} shows that our models do not reach the efficiencies reported by \citet{TenorioTagle1990MNRAS244p563}. This is partly caused by our higher ambient density, but most importantly we evaluate the feedback energy efficiency at much later times. 

That our simulations can reproduce the results of \citet{TenorioTagle1990MNRAS244p563} for the same initial conditions is shown in the Appendix in Fig.~\ref{fig:compTenorio}. For illustration, we show the density structure of one model at different evolutionary stages in Fig.~\ref{nh100x200+rho}.

\subsection{Grid of models}
The ambient medium in our simulations has a density of $\rho_0=2.2\times 10^{-22}$~g~\mcmb and is in cooling-heating equilibrium of the cooling-heating prescription. In cells with densities ($\rho$) above  $a \rho_0$ radiative cooling is taken into account (see also section~\ref{sect:separate}). Less dense cells do not suffer cooling losses. The grid of models spans $a=0$ to $1.3$ (here only $a=0$ and $a=1$ are shown) and the resolutions of $1$, $2$, $4$,  $8$ or $16$ cells per $0.064$~pc ($\sim 2\times 10^{17}$~cm). For reference some models do not contain winds or SNe (Table~\ref{tab:newGrid}) or insert the SN in a spatially upsampled wind bubble structure of a lower resolution run.
 
\subsection{Comparison to previous work}
\begin{figure}
\includegraphics[width=\columnwidth, trim = 2mm  2mm 2mm 2mm, clip]{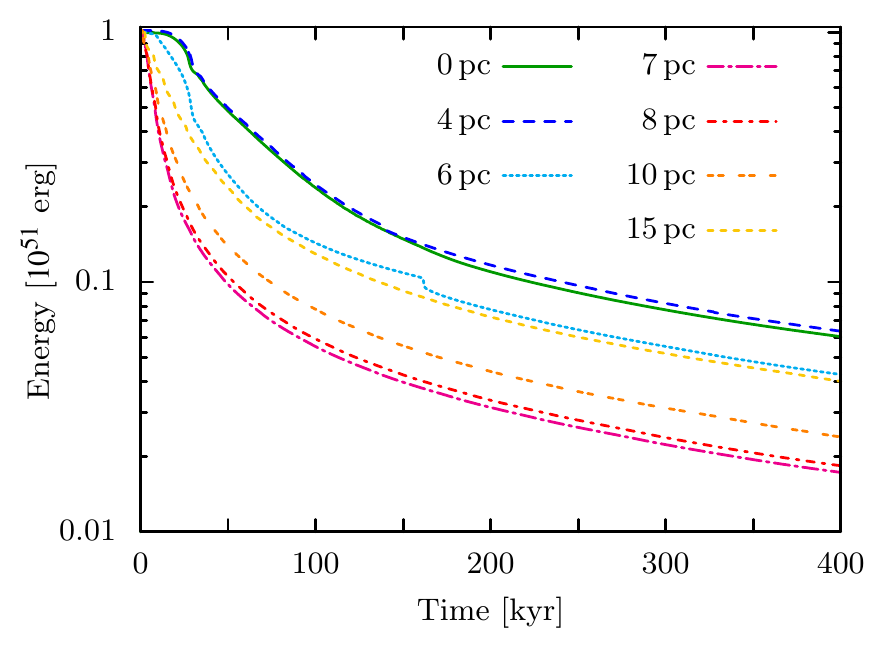}
\caption{Minimal energy bubbles. In this study SNe exploded at $t=0$ in a cavity of given radius. These cavities were created by a constant wind and the ambient medium has $n_{\rm H}=1$~\mcm, $T=1\,000$~K. The feedback efficiency of a SN in a pre-existing bubble depends on the bubble size, since on the one hand, the bubble can act as a pressure reservoir due to the very small cooling losses inside the bubble and on the other hand the dense cavity walls lead to large radiative losses. It can be seen that bubbles of $\sim 7$~pc radius have the smallest feedback energy efficiency. Such bubbles are however, even too small for the winds of the least massive star ending in a SN. For larger radii the feedback efficiency rises with increasing radius.}
\label{fig:energyminimum}
\end{figure}
\begin{figure}
 \includegraphics[width=\columnwidth, trim = 2mm  1mm 1mm 1mm]{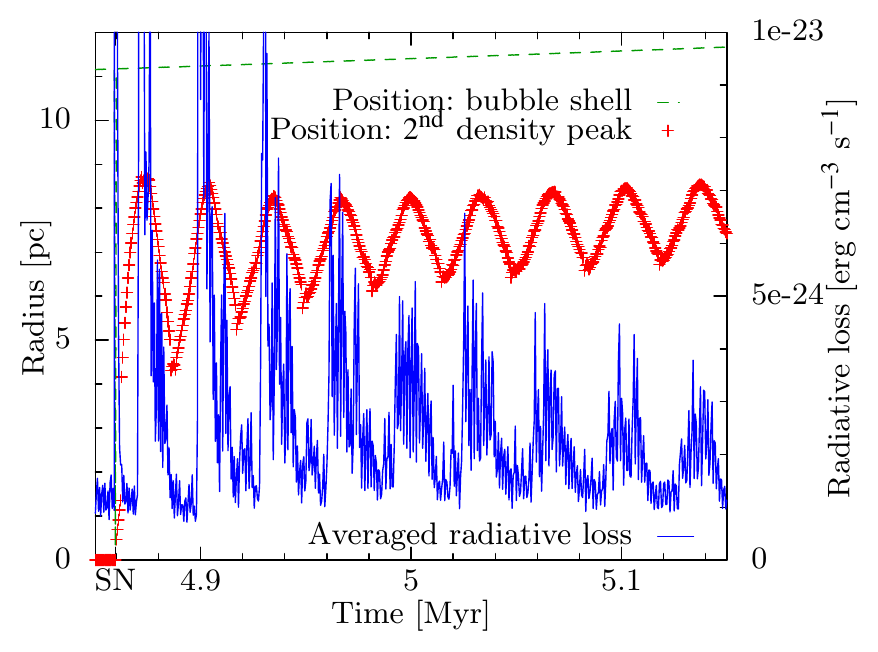}
\caption{Motion of the secondary density peak compared to the evolution of the averaged radiative losses. The losses peak, when the wind material close to the bubble wall gets compressed. This phase also leads to the sudden drop of the retained energy of the $6$~pc bubble in Fig.~\ref{fig:energyminimum} at $\sim 150$~kyr.}
\label{fig:bouncing}
\end{figure}
\begin{figure*}
\includegraphics[width=\textwidth, trim = 2mm  1mm 1mm 1mm, clip]{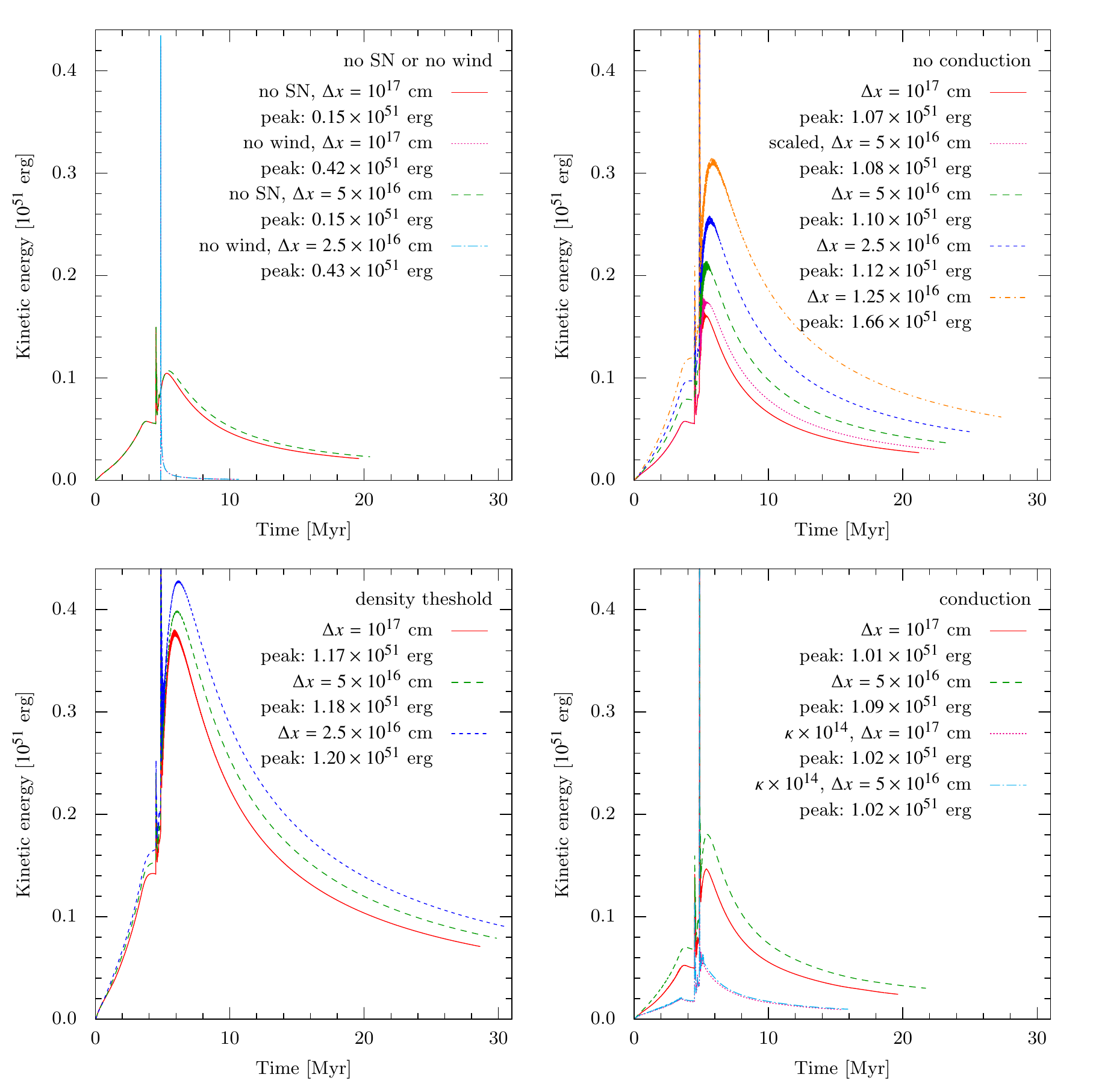}
 \caption{Time evolution of the retained kinetic energy. The wind phase ends after $4.86$~Myr. All lines end when the cell with the highest density is decelerated to the sound speed of the ambient medium. Simulations with a supernova without pre-existing wind bubble (left upper panel) have a six times lower feedback energy efficiency than the more realistic models with supernovae in pre-existing bubbles (right upper panel) [see also Table~\ref{tab:newGrid}: for $\Delta x=0.032$~pc the simulation with a supernova without wind leads to $0.11\times 10^{49}$~erg of kinetic energy compared to the difference between simulations with wind and with/without supernova: $0.58\times 10^{49}$~erg]. $t_{\rm f}=13 t_0$ [$t_0$ is the time of maximal loss, at $t_{\rm f}$ the efficiencies are evaluated] as defined by \citet{Thornton1998ApJ500p95} is $4.8915$~Myr for the model without wind (kinetic shell energy: $0.61\times 10^{50}$~erg) and ranges from $4.9955$~Myr to $5.0605$~Myr for all other simulations. The maximum of the y-axis was chosen to reduce white space and make comparisons between the four panels and Fig.~\ref{fig:newgrid:E+v} easier. The given peak values should not be over-interpreted as the peaks are very transient phenomena. In pre-existing bubbles almost all thermal energy of the SN is converted to kinetic energy, but is quickly lost, when the blast hits the shell.}
 \label{fig:newgrid:E+t}
\end{figure*}
\begin{figure*}
\includegraphics[width=\textwidth, trim = 2mm  1mm 1mm 1mm, clip]{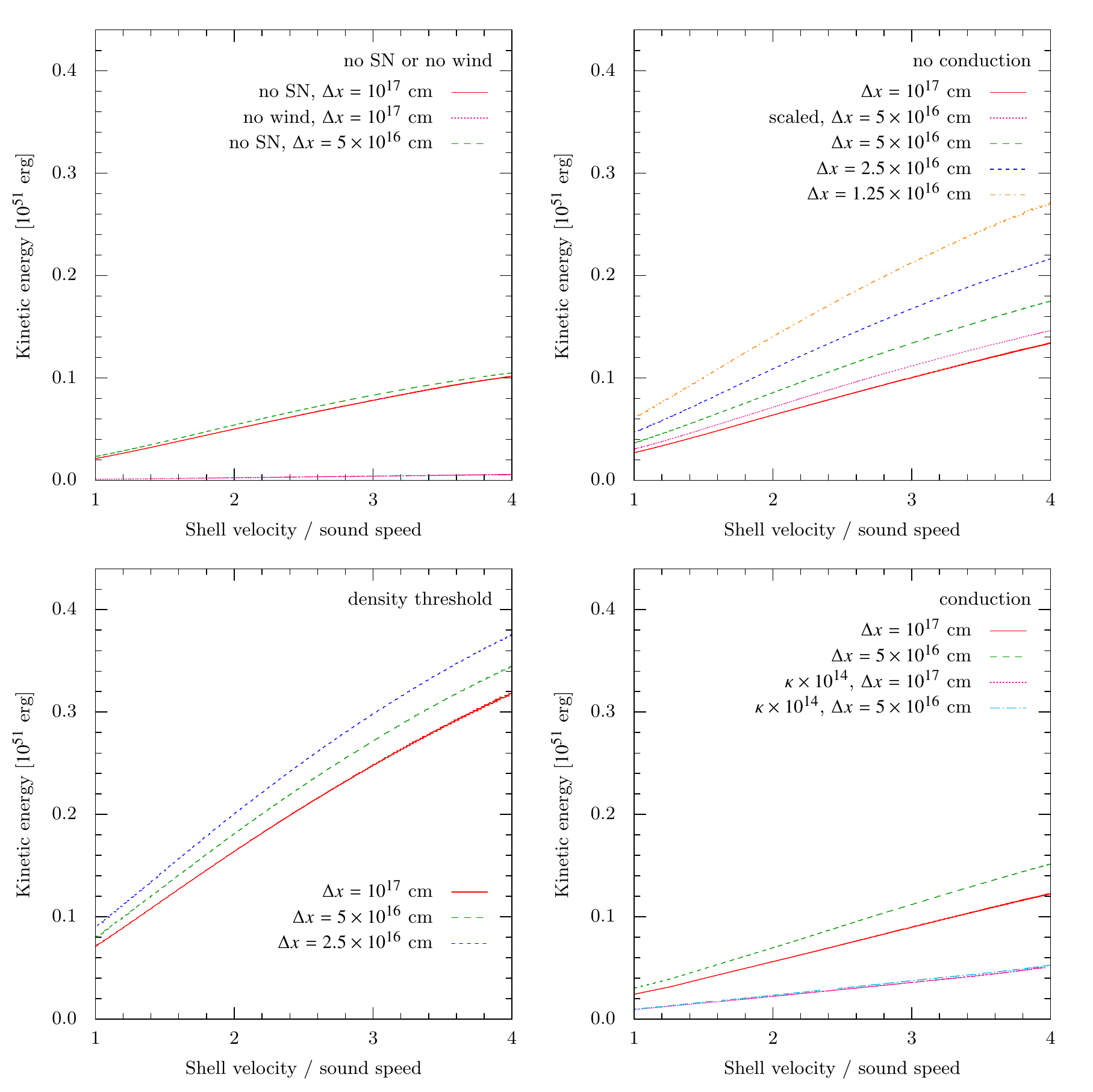}
 \caption{Retained kinetic energy as a function of the velocity of the cell with the highest density. This is a variant of Fig.~\ref{fig:newgrid:E+t}: The retained kinetic energy for the same models is displayed as a function of velocity instead of time. The velocities are normalized to the sound speed of the ambient medium ($\sim 1$~km~s$^{-1}$).} 
 \label{fig:newgrid:E+v}
\end{figure*}

\citet{TenorioTagle1996AJ111p1641} report a dichotomy of wind-blown bubbles (1) light bubbles, which are overrun by the SN-shock and (2) stable bubbles that switch to the radiative phase as soon as they are hit by the blast. 

For reference we produced stellar wind bubbles with a constant wind mimicking a $40$~\Msun star consistent with the feedback in \citet{TenorioTagle1990MNRAS244p563} with a terminal wind velocity of $v=1\,000$~km~s$^{-1}$ and a mass loss rate of $\dot{M}=3 \times 10^{-5}$~\Msun~yr$^{-1}$, immersed it in a $n_{\rm H}=1$~\mcm, $T=1\,000$~K medium and ignited the SN as soon as the desired bubble radius was reached. Fig.~\ref{fig:energyminimum} shows that we observed minimal energy bubbles in between these two cases: The minimal efficiency occurred at ``intermediate'' cavity sizes of $7$~pc in a $n_{\rm H}=1$~\mcm, $T=1\,000$~K medium. This minimum is created by the counteracting effects of efficient cooling in the denser shells of larger bubbles and the larger cavities with inefficient cooling serving as pressure reservoirs. However, this minimum is of academic interest only, since modelling the wind of the lowest mass star that still ends in a SN shows that nature does not produce these minimal energy bubbles: Even a $9$~\Msun star in a $n_{\rm H}=100$~\mcmb medium can produce a $10$~pc cavity before ending in a SN explosion.

\subsubsection{Wind phase}
During the stellar wind phase the models show the structure expected from stellar wind bubble theory \citep{Pikelner1968ApL2p97,Avedisova1972SvA15p708,Castor1975ApJ200p107,Weaver1977ApJ218p377,Dyson1977AA59p161}. Earlier models neglect radiative losses in the shell and describe the evolution of the shell with three parameters: the mechanical energy input ($L=0.5\dot{M}v^2_{\infty}$ with the mass loss rate $\dot{M}$ and terminal wind velocity $v_{\infty}$), the ambient number density ($n$) and the age ($t$). The bubble radius is:
$
R \propto \left(\frac{L}{n}\right)^{2/3} t^{3/5} 
$.
However, the feedback energy efficiency of $100$ per cent (i.e. all feedback energy is retained as thermal or kinetic energy, since no energy losses via radiative cooling are taken into account) postulated by this model is unrealistic. Obviously the extent of the zones and zone width ratios have to differ from these simple adiabatic models, since we allow for radiative losses: For example the swept-up shell is thinner, denser and moves more slowly into the ambient medium.

Before the SN explosion and shortly after, the leftmost part of Fig.~\ref{nh100x200+rho} (below $2$~pc) shows the typical density structure of a free streaming wind. This part of the plot can be compared to the solution of \citet{Chevalier1985Natur317p44}. In our models a region of freely expanding wind, containing cold gas and mostly kinetic energy, is separated from the thermalized ejecta, which consist of hot dilute plasma and contain mostly thermal energy, by a reverse shock. The presence of this free expansion zone in our simulations shows that our feedback region radius is not too large.

The $p{\rm d}V$ work of the thermalized ejecta sweeps up the ambient medium. This medium forms a thin, efficiently cooling shell, which is separated from the thermalized ejecta by a  contact discontinuity (CD). Due to the absence of pressure and velocity gradients across this surface, no mixing (except for diffusion) between the medium inside and outside the CD is expected \citep{TenorioTagle1996AJ111p1641}.
\subsubsection{Post SN phase}
If a SN explodes in the wind bubble of its progenitor, the blast wave expands freely and also adiabatically in the dilute medium inside the wind bubble. 

Thus, in a pre-existing cavity the Sedov expansion phase is skipped \citep[see e.g.][]{TenorioTagle1990MNRAS244p563}, since after this free expansion phase, when the blast wave hits the bubble wall, the evolution continues snowplow-phase-like. In fact, the SN ejecta do not reach the dense shell. They rather compress the wind gas and get reflected (Fig.~\ref{fig:bouncing}). Thus, according to our models, the velocity of the SN-ejecta is expected to be higher than the velocity of the gas in the bubble wall. After the onset of the increased mass loss in the WR phase and after the SN, a bouncing wave inside the cavity is also visible in Fig.~\ref{nh100x200+rho}.

After reflection from the bubble wall the SN blast wave continues to travel back and forth inside the cavity. This causes oscillations in the kinetic and thermal energy evolution as well as in the cooling losses: Whenever the wave hits the wind gas in front of the bubble wall, compresses it and gets reflected, the radiative losses peak. The losses at the conversion from kinetic energy to thermal energy are larger than at the backward-conversion to kinetic energy (see also Fig.~\ref{fig:bouncing} and \ref{fig:newgrid:E+t} as well as \ref{fig:compTenorio} in the Appendix).

 As in the models without progenitor winds the cold outer shell is accelerated by $p{\rm d}V$ work from the hot (SN) gas inside the bubble. In later stages, when the pressure in the bubble becomes ineffective, momentum conservation pushes the shell into the ambient medium. At the end of the pressure driven phase a considerable widening of the shell can be observed in Fig.~\ref{nh100x200+rho}. As a consequence, models with different spatial resolutions can converge during this phase.
\subsection{Feedback energy efficiency: winds or SNe?}
Fig.~\ref{fig:newgrid:E+t} and \ref{fig:newgrid:E+v} show the kinetic energy evolution of our models summarized in Table~\ref{tab:newGrid}. The efficiencies listed in Table~\ref{tab:newGrid} were computed at the moment when the cell with the highest density in the simulation moves slower than the sound speed of the am\-bient medium. In Fig.~\ref{fig:newgrid:E+t}, which shows the time evolution of the retained kinetic energy, the lines also end at this time. Fig.~\ref{fig:newgrid:E+v} depicts the retained kinetic energy of all these models as a function of the shell velocity.

For these models time resolved stellar winds of a $60$~\Msun star were blown into a homogeneous medium. The time of the SN explosion is set by the stellar model, thus the wind bubble size can only be influenced indirectly via the density of the ambient medium and the chosen stellar model. (In contrast to the constant wind test shown in Fig.~\ref{fig:energyminimum}, where the SN explosions occur at a pre-defined bubble size).

To compare the feedback efficiency of winds and SNe, some models in our grid lack the SN explosion or the wind phase. They are shown in the left upper panel of Fig.~\ref{fig:newgrid:E+t} and \ref{fig:newgrid:E+v} and in Table~\ref{tab:newGrid}. Since the $60$~\Msun model explodes in a SN after $4.86$~Myr, models without wind phases are started at this time. 

 The model without SN explosion demonstrates the importance of stellar winds: The total energy input into the wind-only model is $2.34\times 10^{51}$~erg, which is $\sim 70$ per cent of the total energy input of a more realistic model with wind and SN. The kinetic energy of the shell in the wind-only model at the time when it is decelerated to the sound speed of the ambient medium is $79$ per cent of the final energy of the model with a SN blast after the wind phase.

Another indication that continuous energy input is more efficient than blasts is the comparison between the model with a constant wind (CW) and the model with time dependent wind strengths (Table~\ref{tab:newGrid}). For reference the same total wind energy is inserted at a constant rate in the CW model. This steady wind has more power at early times (Fig.~\ref{fig:simplemodel}), since the energy input of the WR phase is distributed over time. Thus, the steady wind produces larger bubbles than a wind with time varying power input, but the same total energy input. Since wind-blown bubbles serve as pressure reservoirs after the SN, higher feedback energy efficiencies are found for larger bubbles.

 Overall it can be seen that wind-bubbles enhance the energy feedback efficiency. For example the models with a resolution of $0.032$~pc without progenitor wind retain $1.1\times 10^{48}$~erg of $10^{51}$~erg ($0.11$ per cent) whereas models with pre-existing bubbles retain more than $4.7\times 10^{49}$~erg of $3.34 \times 10^{51}$~erg ($1.5$ per cent).
\subsection{Convergence of the retained kinetic energy}
We have checked the convergence of our models for different Riemann solvers as well as different temporal and spatial resolutions. Details on the two aforementioned studies can be found in the Appendix (Sect.~\ref{sect:convergence}). 

Generally speaking we found no dependence on the time-step size and increasing the diffusivity of the Riemann solver has similar effects as decreasing the spatial resolution. Our models converge if cooling losses in the newly swept-up medium dominate. This is the case in momentum driven bubbles (i.e. in all our models for SNe without progenitor-winds and at late phases of the other models), whereas our models can not converge when the cooling losses caused by mixing across the CD dominate in the pressure driven bubbles (e.g.\ during the wind phases and in the early post-SN phase). This convergence issue can, however, be solved by deciding on which scales the ISM mixes (see Appendix, Sect.~\ref{sect:conduction} to \ref{sect:turbdiff}). The spatial resolution of the numerical simulation governs the mixing of gas phases across the CD (the {\sc Pluto} code allows for one gas phase per cell) and thus implies a length scale on which diffusive processes occur. Thus, the feedback energy efficiencies of our simulations with different resolutions are solutions for different efficiencies and scale lengths of turbulent diffusion. 

In short, the phase when the maximal velocity in the shell falls below the sound speed of the ambient medium occurs later, at larger radii and at higher kinetic energies for higher $a$ and higher resolutions.
\subsubsection{Spatial resolution}
In contrast to the feedback energy efficiencies of SNe in a homogeneous medium, the efficiencies of SNe in a pre-existing bubble depend on the assumed length scale of mixing in the ISM (Appendix, Sect.~\ref{sect:conduction} to \ref{sect:turbdiff}). If the assumed length scale of the mixing processes is below our resolution, the efficiencies in Table~\ref{tab:newGrid} are lower limits.

Table~\ref{tab:newGrid} and the right upper panel of Fig.~\ref{fig:newgrid:E+t} show that resampling the wind bubble to twice the resolution at the SN leads to an increase of the retained kinetic energy. If the model is resampled to twice the resolution after $6$~Myr, as soon as the oscillations due to the evanescent SN wave, which can be seen e.g.\ in Fig.~\ref{fig:bouncing}, are damped away, also the feedback energy efficiency in the rescaled model is higher. Restarting at the end of the pressure driven phase ($9$~Myr, not shown in the plot, since the lines would be on top of each other) with twice the resolution does not change the efficiency. This is consistent with the SN model without wind, which retained $0.11$ per cent of the inserted energy when the shell speed reached the sound speed of the ambient medium independently of the resolution. $0.2$~Myr after the SN all models without pre-existing bubble have entered the momentum conserving phase (the transition times are listed in Table~\ref{tab:NOwind:PdV} and \ref{tab:NOwind:PdV:shell}).

Basically the cooling losses occur in two distinct regions of the models: in the dense, swept-up shell and near the CD. In simulations with low spatial resolutions\footnote{These models have lower resolutions than the models in Table~\ref{tab:newGrid}.} the swept-up shell is not resolved and the cooling losses in the dense shell dominate. Thus, increasing the resolution reduces the energy efficiency, since it causes higher peak densities in the swept-up shells and the cooling losses rise with number density squared. At higher spatial resolutions, as soon as mixing across the CD produces a strongly cooling cell at every time-step\footnote{A strongly cooling cell arises if enough energy from the hot phase is mixed with enough density from the cold phase. At low resolution this occurs only less frequently (only every n-th time-step).}, however, the feedback energy efficiency starts to rise again with increasing resolution. This behaviour is caused by the volume of the strongly cooling zone\footnote{The  cooling losses are proportional to the volume, the time and the density squared. The density in the mixing cell is independent of the resolution, since the flux of hot gas into the CD cell is set by the shell velocity. The mixture in the cell is given by:
$n_{\rm average} = n_{\rm hot} v_{shell} \frac{\Delta t}{\Delta x} + \left(1 - v_{shell} \frac{\Delta t}{\Delta x}\right) n_{\rm cold}$
or
$n_{\rm average} = \left(n_{\rm hot} - n_{\rm cold} \right) v_{shell} \frac{\Delta t}{\Delta x} + n_{\rm cold}$.
$\frac{\Delta t}{\Delta x}$ is set by the peak velocity and the CFL. This shell velocity to peak velocity ratio differs less than $10$ per cent between the resolutions. Moreover, the density is smaller for smaller CFL, but the in our simulations the energy efficiencies did not depend on the CFL.}: This zone is located at the CD and has a width of a single cell only. The volume decreases, if the cell sizes are reduced. There are two counteracting effects: (1) the volume of a shell with a width of one cell at the same radius is reduced by the factor $\frac{\Delta x_1}{\Delta x_2}$ (i.e. $0.5$ for doubling the cell number), but (2) at the same simulation time, simulations with higher resolution and thus higher efficiency have already produced larger bubbles. This makes the volume ratio at the same simulation time larger than $\frac{\Delta x_1}{\Delta x_2}$ i.e. $>0.5$ for doubling the cell number.

If the energy losses at the CD dominate, we would expect to find half the loss if the volume of the cooling zone is halved. From Table~\ref{tab:newGrid} we find, however, that the kinetic energy of the shell when the shell has been decelerated to the sound speed seems to rise like $E_0 \times (1.3)^n$ for $a=0$ and like $E_0 \times (1.1)^n$ for $a=1$, where $n$ is the number of cells per unit length and $E_0$ is a proportionality constant. The lower factor for $a=1$ strengthens the assumption that this treatment of the CD reduces the importance of radiative losses near the CD in this model.

For the resampled model without SN and $a=0$, we find a factor $\sim 1.1$ despite the fact that no energy is added to this model after resampling. This is expected, since the absence of the SN blast reduces the duration of the pressure driven phase and in our simulations differences due to the spatial resolution arise in pressure driven phases only. The higher resolved model can retain more kinetic energy, since it loses less energy at the CD.

The comparison of these factors and the fact that resampling the model after the transition to the momentum driven phase to higher resolution does not influence the feedback energy efficiency show that the treatment of the CD and the assumed mixing processes are most important during the wind phases and the pressure driven post-SN phase. 

The influence of the spatial interpolation scheme on the retained energy is described in Appendix Sect.~\ref{sect:solver}.
\subsubsection{Influence of the size of the feedback region}
To test the influence of the number of cells in the feedback region on the energy content of the simulation, models with different resolutions ($\Delta x$ from $0.008$~pc to $0.032$~pc) and diameters of the feedback region ($r_{\rm f}$ from $0.32$~pc to $0.64$~pc) were compared.

This set of models shows the general trend that simulations with higher spatial resolution find higher energy efficiencies. Comparing the free streaming region to the solution of \citet{Chevalier1985Natur317p44} shows good agreement for all models: The density profile is $\sim\frac{1}{30 x^2}$ for all $\Delta x$ and all $r_{\rm f}$. Also the kinetic energy profiles for all $\Delta x$ and all $r_{\rm f}$ are similar to those in \citet{Chevalier1985Natur317p44}. Since the pressure in the top hat distribution in the feedback region is proportional to $r_{\rm f}^{-2}$, the pressure is larger for larger $r_{\rm f}$. All models show a decay like $p\propto x^{-10/3}$, as expected.

The kinetic and thermal energy increase starts later for $\Delta x= 0.016$~pc and $r_{\rm f}=0.64$~pc than for $r_{\rm f}=0.32$~pc at the same resolution, since the initial top hat structure has to evolve into a wind structure, which takes longer for larger regions. Later the rate of energy increase is the same. I.e.\ adjusting the zero points of time in the energy vs.\ time diagram shows the same rise. As a result increasing $r_{\rm f}$ leads to slightly smaller bubbles. However, if the spatial resolution is decreased to $\Delta x = 0.032$~pc, the energy increase also starts later for larger $r_{\rm f}$, but after $0.1$~Myr the energy uptake rate becomes higher for larger $r_{\rm f}$, leading to larger bubbles for larger $r_{\rm f}$. Doubling the feedback region radius thus led to an increased energy efficiency for the lowest resolution, since the larger pre-existing bubble can serve better as pressure reservoir. For  $\Delta x = 0.016$~pc however, the region diameter did not change the efficiency any more. 
\subsubsection{Influence of mixing processes}\label{density:threshold}
If radiative cooling is applied for all densities in the cooling table ($a=0$, Table~\ref{tab:newGrid}, Fig.~\ref{fig:newgrid:E+t} right upper panel), the kinetic energy at the end of the wind phase is a factor $1.3$ higher in simulations with a cell size of $\Delta x =0.008$~pc than in simulations with $\Delta x =0.016$~pc. In the latter, the kinetic energy during the wind phase is a factor $1.2$ higher than in a simulation with $\Delta x =0.032$~pc. At the end of the simulations, when the bubble shell has decelerated to the sound speed of the ambient medium, the feedback energy efficiency rises by a factor $1.3$ if the number of cells is doubled.

If there is no density threshold for radiative cooling ($a=0$), also the SN shell can cool. More than $70$ per cent of the energy is lost via radiative cooling when the SN blast hits the bubble wall. All the kinetic energy in the reflected wave is lost at the origin, since the reflected wave sweeps up the gas and creates an efficiently cooling density peak at the origin. Again losses are higher in simulations with larger cells.

In the left lower panel in Fig.~\ref{fig:newgrid:E+t} and \ref{fig:newgrid:E+v} the CD is artificially enforced via the threshold density $a$ for cooling. The dependence on the resolution in these models is less pronounced than in the more realistic cases shown in the right upper panel, but still exists, since the treatment with $a$ reduces the losses near the CD, but cannot prevent mixing of the two phases. Limiting the mixing processes across the CD by applying radiative cooling only to cells with densities above the ambient density, leads to a feedback energy efficiency of approximately $7$ per cent for a cell size of $\Delta x =0.032$~pc. If all cells with densities below the ambient density are considered to contain not radiatively cooling hot gas ($a=1.0$, Table~\ref{tab:newGrid}, Fig.~\ref{fig:newgrid:E+t} left lower panel), halving the cell size increases the kinetic energy when the bubble shell has decelerated to the sound speed of the ambient medium or the kinetic energy at the end of the wind phase by a factor of $1.1$. If the cell size is reduced, the oscillations between kinetic and thermal energy caused by the SN are less damped. The radiative energy losses are largest when thermal energy is converted to kinetic energy (see Fig.~\ref{fig:bouncing}). When the wave enhances the pressure near the bubble wall, strong radiative cooling losses arise in cells, which are dense and hot enough to cool. Since no density peak (as high as the ambient medium) is found at the origin no additional losses occur when the SN wave is reflected at the origin. The losses are larger if the cells are larger. 

The right lower panel shows the second approach to limit the losses near the CD: A mixing process smears out the CD and produces several cells with intermediate temperature gas and intermediate density gas. This prevents that high temperature gas mixes with dense gas at the CD. Taking into account thermal conduction (see Appendix Sect.~\ref{sect:conduction}) lowers the efficiencies by $10$ per cent ($\Delta x = 0.032$~pc) or $18$ per cent ($\Delta x = 0.016$~pc). Also the dependence on the spatial resolution decreases, if thermal conduction is taken into account. In this panel we also show a $14$ orders of magnitude higher diffusion coefficient to mimic a very efficient mixing process. Efficient mixing is expected to remove the dependence of the feedback energy efficiency on resolution, and indeed, the model with extreme conduction is converged for all resolutions.

Basically our spatial resolution defines a scale length on which gases are mixed with $100$ per cent efficiency. Since our resolution has reached or even gone below the proposed length scale of turbulent mixing (Appendix Sect.~\ref{sect:turbdiff}) we conclude that the dependence of the feedback energy efficiency on the spatial resolution depicts the dependence of the radiative losses caused by mixing across the CD.

\section{Discussion}
\begin{figure}
\includegraphics[width=\columnwidth, trim = 2mm 0mm 1mm 1mm, clip]{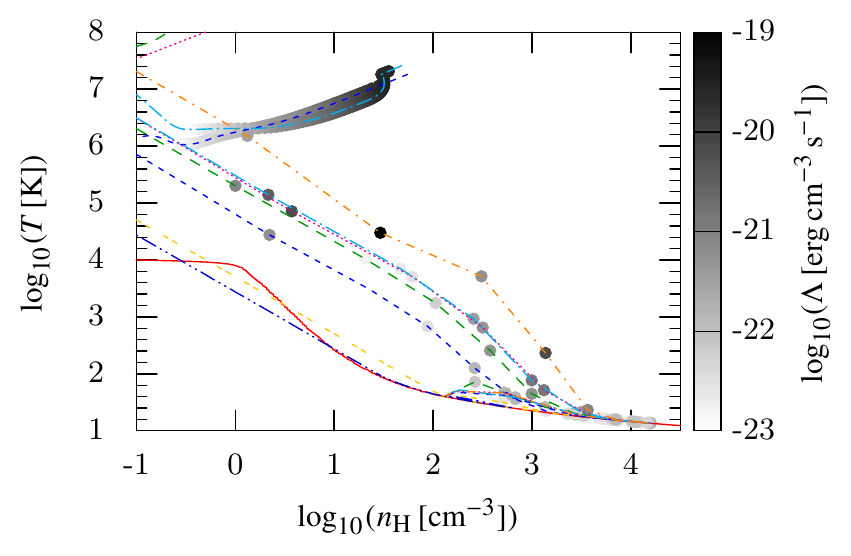}\\
\includegraphics[width=\columnwidth, trim = 2mm  0mm 1mm 1mm, clip]{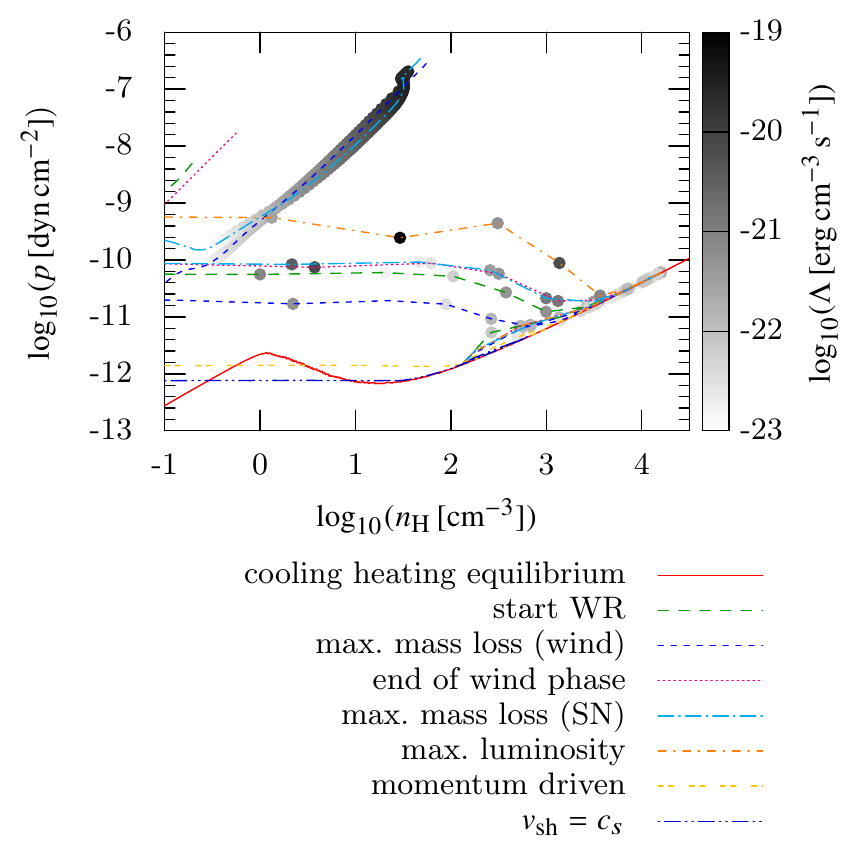}
\caption{Gas phases in the $a=0$, $\Delta x=0.016$~pc model: The cells with the highest density approach the cooling-heating equilibrium (solid line). The fill colour of the dots carries information on the radiative losses. The dark colours of the rightmost points on the curves show cooling losses in the dense, swept-up shell. Bright points on the equilibrium curve depict the ambient medium. The solid lines connecting dots are meant to guide the eye and link gas from adjacent cells. The dark dots in the centre show the cooling losses near the CD. The plot compares the location of the cooling losses at different stages of the evolution. When the mass loss rate peaks, cooling losses of dense gas are found near the feedback zone. Typical pressure driven phases (start of the WR phase at $3.46$~Myr, end of the wind phase at $4.50$~Myr, time of maximal luminosity at $4.87$~Myr) show cooling losses near the CD whereas losses in the dense shell dominate during momentum driven phases (in the plot the start of the momentum driven phase at $9.76$~Myr and the end of the simulation at $23.40$~Myr are shown).}
\label{eq+t+nh100x200}
\end{figure}
As we have shown, blast waves of SN explosions in cavities excavated by WR winds undergo an expansion almost without energy losses until they hit the cavity walls. As a consequence wind-blown bubbles delay the time of maximal luminosity (defined in Sect.~\ref{sect:feedbackmodel}) and increase the amount of retained energy, since such cavities can act as pressure reservoirs. When the blast hits the cavity walls, so-called catastrophic cooling in the dense shell of swept-up ambient medium sets in \citep{TenorioTagle1990MNRAS244p563,Smith2003MNRAS339p133}. This evolutionary stage should be observable, and it has been suggested \citep{Chu1990ApJ365p510,Arthur1996ApJ457p752,Oey1996ApJ467p666} that the X-ray emission in excess of an adiabatic model in X-ray bright super-bubbles is likely caused by a SN blast wave hitting a pre-existing shell and leading to strong radiative cooling losses.

Since the feedback energy efficiency is greatly influenced by a limited number of cells suffering strong radiative losses, we will briefly summarize the nature of these cells. The largest cooling losses of the models are
\begin{itemize}
\item at the CD during pressure driven phases.
\item in the dense shell during momentum conserving phases.
\end{itemize}
High resolution simulations have a higher feedback-efficiency during the wind phase (and other pressure driven phases) because
\begin{itemize}
\item the volume of the strongly cooling layer gets smaller at higher resolutions. 
However, at very low resolutions the feedback-efficiency starts to rise again, since in this case a cell suffering from large radiative losses caused by mixing across the CD does not exist at every time step, because the gas does not get dense or hot enough to cool efficiently.
\item smaller cells lead to a better separation of the media. Decreasing the cell size can thus mimic a gas with less efficient mixing processes (physical processes are discussed in Appendix~\ref{CDdiffusion}).
\end{itemize}

The deviations from the cooling-heating equilibrium and the cooling losses are shown in Fig.~\ref{eq+t+nh100x200}. In this figure the evolution of the gas phases in the $a=0$, $\Delta x=0.016$~pc model are visualized. The solid line shows the cooling-heating equilibrium curve. The ambient medium is represented by a very bright dot (no losses) on the equilibrium curve. The gas properties in the swept-up shell and inside the bubble are shown by dots linked with lines connecting adjacent cells. The colour of the dots contains information on the radiative losses. It can be seen that there are two regions with enhanced cooling losses: the CD (centre) and the dense part of the shell (bottom right). The cooling-heating phase space plot shows seven distinct snapshots of the model represented by different line styles:
 (1) $3.457$~Myr at the start of the WR phase the shell is pressure driven and we find cooling losses near the CD and in the shell, 
 (2) the mass loss rate of the winds peaks at $4.4975$~Myr and leads to dense, cooling gas near the feedback region.
 (3) Towards the end of the wind phase at $4.85$~Myr radiative cooling is effective in the shell and near the CD. 
 (4) As soon as the SN explosion has taken place ($4.8596$~Myr), again dense material is found near the feedback region.
 (5) At the time of maximal luminosity ($4.8695$~Myr), when the SN blast wave hits the cavity wall, cooling near the CD is very important.
 (6) When the model has transited to the momentum driven phase ($9.7595$~Myr) cooling in the dense, swept-up shell dominates. At this stage models of different spatial resolution converge. 
 (7) Also at the end of the simulation ($23.3975$~Myr), when the shell has decelerated to the sound speed of the ambient medium, cooling is only effective in the dense shell.
 
Comparing cooling losses in these snapshots shows that the energy losses in or near the CD cell are less important after the end of the pressure driven phase. At this point the models of different spatial resolution start to converge. 

In short, wind-blown cavities should not be ignored as they can strongly increase the amount of kinetic energy deposited in the ambient medium by reducing radiative cooling losses. For the same reason it is dangerous to argue that one can safely ignore the feedback of the most massive stars (O stars), as they are relatively rare and rather focus on the more abundant B stars, which also end in a SN explosion. Just like not taking stellar winds into account, this approach misses the important effect, the density structure of the ambient medium plays in determining how efficiently the SN energy can be converted to kinetic energy of the ambient medium. Stellar winds are often assumed to be constant if they are taken into account in the literature on feedback energy efficiencies \citep[e.g.][]{TenorioTagle1990MNRAS244p563,TenorioTagle1991MNRAS251p318,TenorioTagle1996AJ111p1641}. We already mentioned that ignoring winds is problematic, since the amount of mechanical luminosity that can be converted to shell motion differs between models, which insert all energy in a blast (a SN) and models where stellar winds are energy sources over long periods of time \citep[see also][and Table~\ref{tab:newGrid}]{TenorioTagle1990MNRAS244p563,TenorioTagle1991MNRAS251p318,Oey1994ApJ425p635,Oey1996ApJ467p666,TenorioTagle1996AJ111p1641}. To a smaller extent one can also run into the same problem, if the time dependence of the wind strength is ignored.

To summarize, during all momentum driven phases, the models converge nicely. However, models with pre-existing bubbles also exhibit pressure driven phases during their evolution. In these phases the mixing of gas phases across the CD leads to non-convergence. Our suggested work around for this problem is to use Fig.~\ref{fig:newgrid:E+t} or \ref{fig:newgrid:E+v}: 
(1) First one needs to select a length scale for mixing. 
(2) Then one selects the simulation with a resolution close to this limit. 
(3) Additionally one decides at which time or velocity one needs the feedback energy efficiency. 
(4) Our simulations were continued until the peak of the shell velocity falls below the sound speed in the $n=100$~\mcm, $T\sim 100$~K medium ($1$~km~s$^{-1}$). E.g. if one decides that the typical turbulent velocity dispersion of a GMC is higher than this and an earlier end of the calculation is needed, Fig.~\ref{fig:newgrid:E+v} can be used to retrieve the feedback energy efficiency at higher shell peak velocities. For the feedback energy efficiency at earlier times Fig.~\ref{fig:newgrid:E+t} can be used.

We recommend to use the models in the right upper panels in Fig.~\ref{fig:newgrid:E+t} or \ref{fig:newgrid:E+v}. All other panels contain models where some processes were modified for comparison. For example the need of increasing conduction by $14$ orders of magnitude to reach convergence shows, that we have to assume a certain mixing efficiency and mixing scale of gas phases to tackle the problem without convergence.

\section{Conclusions}
We investigated the efficiency of stellar energy deposition in the ISM from massive stars and their SNe in different environments. Our results are:
\begin{itemize}
\item If a simulation with $100$~\pccmb uses a feedback energy efficiency of $10$ per cent as a sub-grid model \citep[as found by][]{Thornton1998ApJ500p95}, a time-step of $33$~kyr (corresponding to $13\,t_0$, the time at which this feedback energy efficiency is found) has to be resolved. A short time later the efficiency drops far below $10$ per cent (Fig.~\ref{fig:NOwind:Ek+t} and \ref{fig:NOwind:Ek+t:c}).
\item Without the stellar wind of the progenitor star the feedback energy efficiency of the SN explosion of a massive star, which is placed in a dense medium, is much smaller than if the wind is taken into account. Table~\ref{tab:newGrid} shows that the retained energies in these two cases differ by a factor $6$.
\item The cumulative feedback energy of the stellar wind of a $60$~\Msun star is $2.34\,E_{\rm SN}$. The impact of the stellar wind can be seen from a comparison between a model with no SN blast at the end of the wind phase and a model with both progenitor wind and SN blast. The energy difference when the shell reaches the sound speed (Table~\ref{tab:newGrid}) is $2.13\times 10^{49}$~erg in a model without SN compared to $2.71\times 10^{49}$~erg in a model with SN and wind. This differs from the ratio of the total energy inputs ($2.34\times 10^{51}$~erg and $3.34\times 10^{51}$~erg). Thus, steady feedback is more efficient than a blast.
\item Models, in which the same net energy input as in the wind with time-varying power input is inserted via a constant wind, show an $8$ per cent higher feedback energy efficiency than models with time resolved winds (see Table~\ref{tab:newGrid}). Averaging the WR phase over the whole stellar lifetime makes the constant wind stronger than the wind with time-varying strength in early phases and allows it to create a larger bubble at early times, which serves as a pressure reservoir for the bubble expansion later on. At the SN the bubble size and the retained kinetic energy of the constant wind model are larger than in the model with varying wind strength whereas the thermal energy is smaller, since the time-varying wind power models boost the thermal energy during the WR phase directly before the SN.
\item The time of maximal luminosity ($t_0$, as defined in \citet{Thornton1998ApJ500p95}) occurs later, if stellar wind bubbles are taken into account. In this case the blast expands adiabatically until it impacts on to the cavity wall. Subsequently the SN blast wave bounces inside the bubble and as a result the luminosity peaks periodically whenever the SN shock-wave hits the cavity wall\footnote{More precisely, it does not directly hit the cavity wall, but compress the wind gas in front of the cavity wall.} and kinetic energy is converted to thermal energy (and vice versa). The losses show a double peak at times when the conversion rates are largest.
\item Mixing processes across the CD are important during pressure driven phases. In these phases the resolution mimics the scale of mixing and thus has an effect on the feedback energy efficiency. Estimates of the physical scale of such mixing processes are discussed in Appendix~\ref{CDdiffusion}. We find that for our setup the estimated length scale on which turbulent diffusion acts ($\sim 1-0.01$~pc) is largest. At the CD in our problem turbulent diffusion seems to be more important than thermal conduction, molecular diffusion and ambipolar diffusion.
 In the subsequent momentum driven phase radiative cooling in the swept-up, compressed and thus heated medium is the dominant energy sink.
\item Comparing the constant wind models at different resolutions (which mimic the length scale of the mixing processes in the ISM) shows that the $0.032$~pc model has a higher efficiency than expected. Low resolution models also can find a higher efficiency, if they underestimate the density in the shell. In this case the efficiently cooling temperature-density combination is not found at every time-step in the $0.032$~pc model whereas later on this phase is always present. In models with higher resolution the efficiently cooling layer near the contact discontinuity has a smaller volume: At the same time of the simulation it is found at larger radii in simulations with higher resolution, but it is only a single cell wide. Simulations with a resolution of $0.001$~pc showed cooling losses of the same order of magnitude in the compressed swept-up medium and near the CD. At even higher resolutions the cooling layer will at some point become irrelevant. 
\item During the wind phase the density threshold in the cooling function (e.g.\ $a=1$) reduces the dependence of the feedback energy efficiency on the resolution (Table~\ref{tab:newGrid}). However, the differences between the feedback energy efficiencies for different resolutions at the end of the simulations are not significantly reduced if the threshold $a=1$ is used instead of $a=0$.
\item If the coefficient $\kappa$ of heat conduction is strongly increased to mimic a highly diffusive process, the models converge, since the gradients at the CD, which were sensitive to spatial resolution, get smeared out. However, the total feedback energy efficiency is drastically lowered by this treatment.
\end{itemize} 
To summarize, winds of massive stars and the cavities created by them, have an important influence on how much of the stellar feedback energy can be used for the ISM dynamics. Since the radiative losses peak near the contact discontinuity, it is necessary to identify the most important process for the degradation of this discontinuity. For example, if turbulent diffusion would act on length scales of approximately $10^{16}$~cm and the mean density of the GMC is $100$~\pccm, it would be possible to convert about $2$ per cent of the stellar feedback energy to kinetic energy of cold gas. This is a lower fraction than the $10$ per cent found by \citet{Thornton1998ApJ500p95} and we evaluate the retained energy at a different phase of the  evolution, but since the stellar wind also provides $2.34\times 10^{51}$~erg, the net energy input ($6\times 10^{49}$~erg) is again of similar order. 

The feedback energy efficiencies from the 1D simulations presented in this work are most likely an upper limit for multi-dimensional simulations, since (non-radial) instabilities, which arise in more dimensions increase the surface of the CD and can thus enhance mixing between the hot and cold gas phase. In our work these mixing processes are treated indirectly via the mixing length-scale (i.e. numerically by the resolution mimicking turbulent diffusion).

\section*{Acknowledgements}
The authors thank the referee for useful comments and suggestions. Also useful discussions with D.~H.~Hartmann and the CAST group are gratefully acknowledged. KF wants to thank A.~Wei{\ss} and the research group of P.~Fierlinger for technical support and computation time. This research was supported by the DFG Cluster of Excellence ``Origin and Structure of the Universe'' and by a Max Planck Fellowship of AB. EN is supported by the ERC Grant Agreement ``ORISTARS'', no. 291294. Additionally this work was supported by funding from Deutsche Forschungsgemeinschaft under DFG project number PR 569/10-1 in the context of the Priority Program 1573 ``Physics of the Interstellar Medium''. MK and MS were partly supported by the Deutsche Forschungsgemeinschaft Priority Program 1573 (``Physics of the Interstellar Medium'').








\appendix

\section{Mean free path}\label{sect:mfp}
A crucial length scale for diffusive processes is the mean free path, which denotes the average distance a particle travels between two scatterings. Processes at the scales of the mean free path and below have to be modelled taking plasma physics into account. Hydrodynamic simulations are based on the fluid approach, which assumes that the mean free path is much smaller than a cell size. Hence we cannot tackle the problem by increasing the resolution, since the underlying assumptions of our method imply a maximal 'meaningful' resolution. The mean free path 
\begin{eqnarray}
\lambda=\frac{1}{\sigma n}\label{eq:lambda:H}
\end{eqnarray}
for elastic scattering of neutral hydrogen with an elastic collision cross section $\sigma_{\rm H-H}$ of $5.7 \times 10^{-15}$~cm$^2$ \citep{Godard2009AA495p847} becomes larger than a cell size of e.g.\ $0.01$~pc (turbulent diffusion length scale estimate of \citet{Gounelle2009ApJ694p1}) if the density falls below $10^{-26}$~g~\mcm, which corresponds to a number density of $0.006$~\mcm. With the mean molecular velocity 
\begin{eqnarray}
v^2_{\rm rms}=\frac{3kT}{m_{\rm H}}=\frac{3RT}{\mu_{\rm mol}} \label{eq:v:rms}
\end{eqnarray}
the average time between collisions is 
\begin{eqnarray}
\tau=\sqrt{\frac{m_{\rm H}}{3kT\sigma^2 n^2}}\qquad . \nonumber 
\end{eqnarray}

In ionized gases the scattering cross section is the area in which the electrostatic energy becomes comparable to the relative kinetic energy of the two charged particles. The electron mean free path 
\begin{eqnarray}
\lambda_e=\frac{0.290\left(kT_e\right)^2}{n_ee^4\ln\Lambda}\nonumber
\end{eqnarray}
 \citep[eq.~5-26][eq.~1.5]{Spitzer1956,Shubook} with the thermal velocity of the electrons 
\begin{eqnarray}
v^2_{T_e}=\frac{kT_e}{m_e}\nonumber
\end{eqnarray}
 and the Coulomb logarithm 
\begin{eqnarray}
\Lambda=\frac{3}{2e^3}\sqrt{\frac{k^3T_e^3}{\pi n_e}}\nonumber
\end{eqnarray}
 is larger than $0.01$~pc for temperatures above $10^{5.36}$~K for densities below $10^{-26}$~g~\mcm.

\section{Diffusive processes in the ISM}\label{CDdiffusion}
The degradation of a contact discontinuity (CD) in kinetic gas theory is caused by particle motion smearing out a gradient. We can look at different manifestations of this microscopic diffusion process. To do so, we think of two distinct gas phases in pressure equilibrium that are separated by a CD. First we will estimate in the rest frame of the CD how many hot particles will flow into an adjacent cold gas and vice versa. This ultimately leads to heat conduction down a temperature gradient (Sect.~\ref{sect:conduction}). Another manifestation of such mixing processes is molecular diffusion (Sect.~\ref{sect:diffusion}). In this case the CD separates two different gas species and diffusion will try to level a concentration gradient. Taking a step back from the microscopic level to the macroscopic level, gas blobs can mix via turbulent diffusion (Sect.~\ref{sect:turbdiff}). And last but not least one can rely on ambipolar diffusion caused by magnetic fields.

\subsection{Evaporation due to thermal conduction}\label{sect:conduction}

In the {\sc Pluto} code \citep{Mignone2007ApJS170p228} thermal evaporation is facilitated with an additional divergence term for heat conduction in the energy equation:
\begin{eqnarray}
\frac{\partial E}{\partial t}&+&\vec{\nabla} \cdot \left[ \left( E + p_t \right) \vec{v}  \right]=-\vec{\nabla} \cdot \vec{F}_{\rm c}  \qquad.\nonumber
\end{eqnarray}

Due to the inverse dependence on the particle mass (evident from the mean molecular velocity, eq.~\ref{eq:v:rms}), conduction is electron dominated. If the scale length of the temperature gradient 
\begin{eqnarray}
l_{\rm T} \equiv \frac{T_{\rm e}}{\left| \nabla T_{\rm e} \right|} \nonumber
\end{eqnarray}
 is much larger than the mean free path of the electrons ($\lambda_{\rm e}$), the heat flux conducting heat down the electron temperature gradient in a plasma is given by 
\begin{eqnarray}
F_{\rm c}=-\kappa_{\rm c}\nabla T_{\rm e} \qquad. \nonumber
\end{eqnarray} 
We use a thermal conduction coefficient for a hydrogen plasma of $\kappa_{\rm c} = 5.6\times 10^{-7} T^{5/2}$~erg~s$^{-1}$~cm$^{-1}$~K$^{-1}$ \citep{Spitzer1962} within the {\sc Pluto} code \citep{Mignone2007ApJS170p228}. 
The relaxation time 
\begin{eqnarray}
t_{\rm relax}=\frac{nc_v}{\kappa_{\rm c}} (\Delta x)^2 = \frac{(\Delta x)^2}{D} = \frac{3}{\bar{v}\lambda} (\Delta x)^2 \nonumber
\end{eqnarray} 
 describes how fast heat conduction in the classic heat flux is. For a gas with a density of $10^{-26}$~g~\mcmb  and a temperature of $10^6$~K on the scales of $\Delta x =0.01$~pc the relaxation time is $\sim 1.8 \times 10^{7}$ years. For steep temperature gradients with scales shorter than the mean free path the code switches to the saturated heat flux, estimated to be 
\begin{eqnarray}
F_{\rm sat}=5\phi\rho c_{\rm s, iso}^3 \,{\rm [erg\,s^{-1}\,cm^{-2}]} \qquad, \nonumber
\end{eqnarray} 
with $\phi=0.3$ \citep{Balbus1982ApJ252p529} and $c_{\rm s, iso}^2=kT/m$ because in this regime the classic heat flux equation overestimates conduction. In the case of a CD we expect such a very steep temperature gradient. For a hydrogen gas with $\rho=10^{-26}$~g~\mcmb and $T=10^6$~K  this flux is $ 1.1\times 10^{-20}$~erg~s$^{-1}$~cm$^{-2}$, which can be compared to the loss via radiative cooling $\Lambda \sim 10^{-22}$~erg~s$^{-1}$~cm$^3n^2 =  10^{-26}$~erg~s$^{-1}$~\mcmb of a slab with a width of $10^{6}$~cm, which is way below our maximal resolution. The heat flux is thus not an important agent near the CD in this problem. 

In our simulations thermal conduction saturated near the CD. The kinetic energy efficiency is only slightly lowered, if thermal conduction is taken into account (Table~\ref{tab:newGrid}, Fig.~\ref{fig:newgrid:E+t}), which is in agreement with the aforementioned order of magnitude estimates.

A more important aspect is the change in particle density, which affects the radiative cooling losses. \citet{TenorioTagle1996AJ111p1641} find $10$ per cent of shell mass mixed into the cavity due to thermal evaporation. The efficiency of mixing of particles of different temperature is discussed in the section on molecular diffusion.

\subsection{Molecular diffusion}\label{sect:diffusion}
Molecular diffusion levels concentration gradients. If a diaphragm between two gaseous species in pressure equilibrium is removed, random movement of all gas particles starts to mix the two species. This process is described with the diffusion equation 
\begin{eqnarray}
\frac{\partial n}{\partial t}=D\frac{\partial^2 n}{\partial x^2} \nonumber
\end{eqnarray} 
with the solution 
\begin{eqnarray}
 n(x,t) = \frac{N}{\sqrt{4\pi Dt}}\exp{\left(-x^2/4Dt\right)} \qquad .\nonumber
\end{eqnarray}  
The diffusion coefficient $D\sim\bar{v}\lambda/3$ is the same as for heat conduction. The diffusion length 
\begin{eqnarray}
\Delta x=\sqrt{2 D t} \sim\sqrt{2/3 \bar{v} \lambda t}\nonumber
\end{eqnarray}  
 is a measure over which physical scales mixing has occurred.\footnote{Diffusion over a scale $\Delta x$ can be found from $(\Delta x)^2 = \overline{\left(x-\bar{x}\right)^2}=\overline{\left(x-0\right)^2}=\frac{1}{N}\int_{-\infty}^{\infty}x^2 n(x,t) dx = 2 D t$.}

This relation can also be used to estimate the time-scale of this process:
\begin{eqnarray}
t_{\rm d}=\frac{\left(\Delta x\right)^2}{\lambda v_{\rm rms}} \label{eq:diff:molecular}
\end{eqnarray} 
with the mean free path $\lambda$ (eq.~\ref{eq:lambda:H}) and the rms-velocity $v_{\rm rms}$ (eq.~\ref {eq:v:rms}).

Equation~\ref{eq:diff:molecular} shows that molecular diffusion mixes chemical species efficiently in the hot dilute gas inside the bubble: In a gas with $n = 10^{-2}$~\mcm, $T=10^{7}$~K and $\mu \sim 1$~g~mol$^{-1}$ we find $v_{\rm rms}\sim 500$~km~s$^{-1}$ and a time of $\sim 33$~years for mixing on the scales of $\Delta x=0.01$~pc.

Diffusion inside the swept-up medium is inefficient ($n = 1$~\mcmb and $T=100$~K leads to a time of $\sim 1.5$~Myr for mixing on the scales of $\Delta x=0.01$~pc).

All particles within a mean free path from the CD can penetrate into the other phase and one sixth of them will have a velocity vector appropriate to do so.\footnote{The number of particles crossing the CD in the time interval $t$ are thus a sixth of the particles within the volume $A v t$ where $A$ is the unit area.} For two phases with $n=0.01$~\mcm, $T=10^6$~K and $n=1$~\mcm, $T=100$~K, respectively, the same number of hot and cold particles cross the CD. There is no change in density and thus no change in the mean free path, but there is a change in temperature. The hot particles in the cold medium undergo their first collision with cold particles after $t=\lambda_{\rm cold}/v_{\rm hot}=0.35$~yr. This means that after $0.35$~years a region of a length of $6\times 10^{-5}$~pc ($\lambda_{\rm cold}$) has a mean temperature of $T_{\rm hot}/6+ 5T_{\rm cold}/6=1.7\times 10^5$~K. To estimate how much thermal energy has been carried into the cold medium we find the number of diffused particles from $\Delta  n = A \lambda_{\rm cold} n_{\rm hot}/6=2.9\times 10^{11}A$~cm$^{-2}$ (with $n_{\rm hot}=0.01$~\mcm, $\lambda_{\rm cold}=1.7\times 10^{14}$~cm). The energy transfer caused by particle motion is $\dot{E}=\dot{n}kT=n_{\rm hot}/6 v_{\rm hot} kT_{\rm hot}=3.6\times 10^{-6}\,{\rm erg\, s^{-1}\, cm^{-2}}$. With a cooling rate of $\Lambda_{\rm cool}=10^{-22}$~erg~s$^{-1}$~cm$^{3}n^2$ the energy flowing through an area $A$ of the CD would be lost in a cell with a number density of $1$~\mcmb and a volume of $A\times 0.01$~pc.

\citet{TenorioTagle1996AJ111p1641} reports that $10$ per cent of the ambient medium ended up in the bubble via thermal conduction and dense clumplets originating from the ambient medium penetrating the bubble wall. From kinetic gas theory we would expect that in each collision time a sixth of the density in the first mean free path of the shell is lost into the bubble. In the example given above, the particle number was conserved, but if the density of the shell is enhanced, there will be a net flux of particles into the cavity.  
\subsection{Turbulent diffusion}\label{sect:turbdiff}

Velocity fields that may be created by hydrodynamic instabilities, overstability of radiative shocks \citep{Chevalier1982ApJ261p543}, non-linear thin shell instability \citep[][NTSI]{Vishniac1994ApJ428p186NTSI}, turbulence or convection can produce eddies and large scale perturbations that are mixed into a different gas phase. Such mixing processes do not necessarily lead to a homogeneous mixture - some authors \citep[for a summary see][]{Pan2012ApJ756p102} rather expect an oil-in-water-like process leading to cold clumps immersed in hot zones, whereas other authors assume that the phases fully mix \citep[e.g.][]{Gounelle2009ApJ694p1}.

In this process eddies of size $l_{\rm turb}$ mix with the velocity $v_{\rm turb}$. The diffusion coefficient of turbulent mixing  is 
\begin{eqnarray}
D_{\rm turb}=v_{\rm turb}l_{\rm turb} \qquad. \nonumber
\end{eqnarray}  
Diffusion rises linearly below the size of turbulent eddies and saturates due to turn-over as soon as the eddy size is reached.

The assumed efficiencies of mixing in a SN shell range from a few per cent \citep[mixing via clumps and RT fingers]{Boss2012ApJ756p9}, over a range from $2$ to $70$ per cent  \citep{Gounelle2012AA545p4}, to the full range of few per cent to full mixing in the study of \citet[clumplets and turbulent diffusion]{Pan2012ApJ756p102}. 

The estimates for the eddy size range from $l_{\rm turb}\sim 0.1-1$~pc \citep[][dispersion of metal-rich droplets in {\sc H\,ii} regions via molecular diffusion and turbulent mixing]{Stasinska2007AA471p193} to  $l_{\rm turb}\sim 0.01$~pc \citep[][highly turbulent mixing process with $100$ per cent mixing efficiency and the characteristic length-scale of the thermal instability]{Gounelle2009ApJ694p1}. Turbulent diffusion is thus likely to act on length scales comparable to the resolution of our simulations.

\subsection{Ambipolar diffusion}
Ambipolar diffusion is a process that can remove magnetic fields from molecular clouds: The magnetic fields are tied to the ionized gas component, and this component drifts relative to the cold, neutral component of the gas, which is accelerated by gravity. E.g.\ \citet{Jijina1999ApJS125p161} noted that ambipolar diffusion takes place more rapidly than the simple laminar description predicts. For a dense core with the size $r$ the time-scale for ambipolar diffusion is $\tau_{\rm AD} = \frac{r}{v_{\rm D}}$ (where $v_{\rm D}$ is the ion-neutral drift speed). \citep[][eq. 81]{1987ppic.proc..491M}
This can be approximated by $\tau_{\rm AD} \sim 3\times 10^6\,{\rm yr}
\left(\frac{n_{\rm H_2}}{10^4\,{\rm cm^{-3}}}\right)^{1.5}
\left(\frac{30\,{\rm\mu G}}{B}\right)^2
\left(\frac{r}{0.1\,{\rm pc}}\right)^2$.

For a density of $1$~\mcmb and a magnetic field strength of $10$~${\rm\mu}$G \citep{Crutcher2012ARA+A50p29} this leads to a time of about three months for $0.01$~pc. However, this process rather acts to separate the gas phases than to mix them.
\subsection{Artificial mixing across the contact discontinuity}\label{sect:separate}
Numerical simulations find large radiative cooling losses near the contact discontinuity (CD) separating the dilute, extremely hot shocked wind gas and the dense swept-up medium. In the literature this is sometimes called ``catastrophic cooling'' \citep{TenorioTagle1990MNRAS244p563,Smith2003MNRAS339p133}. These losses arise because the code mixes two media that should be separated by a CD and the cell with the mixture of the two phases efficiently cools, acting like a valve, considerably reducing the feedback efficiency. If the mixing scale is not resolved numerically, this process could lead to artificially high radiative losses.

In this work we also test the importance of this effect by regulating the radiative energy loss of the critical cell near the CD, which acts as the dominant energy sink. Numerically there are basically two strategies to prevent extreme cooling losses in cells in the vicinity of a CD where the two media mix:

(1) Strictly enforcing the separation of these two phases: The simplest way to avoid cooling losses in the hot, dilute cells in which shell material and wind material can be found, is to increase the density threshold of the cooling function. Our cooling function is tabulated for number densities $n_{\rm H} > 0.01$~\mcm. To avoid cooling losses at the CD, in the models with ``density thresholds'' radiative cooling is switched off if the cell's density is below  $a$ times the ambient density $\rho_0$. For example in runs with $a=1$ radiative cooling is switched off at all densities below the ambient density. By doing this, we mimic a sub-grid model with two nicely separated ISM components in the cell: The gas is either too cold or not dense enough to cool and no strongly cooling intermediate phase is produced. Or in other words, at densities below $a\rho_0$ the simulation becomes adiabatic. \citet{Krumholz2007ApJ671p518} discuss a similar solution to avoid artificially high radiative cooling rates near ionization fronts. Their zone selection is based on the ionization degree instead of the density (as used in our approach).

(2) Postulating a strong mixing process that smears out the temperature and density gradients near the CD: This leads to low temperatures in regions, which are dense enough to cool. Efficiently mixing gas across the CD can be achieved e.g.\ via turbulent diffusion, as discussed before. The radiative cooling losses are a function of temperature and density. Lowering the density and the temperature by enhancing the mixing at the discontinuity can limit the energy losses via radiative cooling by producing cells that are already too cold to cool efficiently. 

\section{Phases of SN bubble evolution}\label{sect:SNevolution}
For a SN explosion in a homogeneous $n=100$~\mcmb ambient medium the initial free expansion phase  quickly transits into the Sedov-Taylor phase ($r\propto t^{2/5}$, $v\propto t^{-3/5}$). This phase ends when the cooling time becomes comparable to the dynamical time. In the subsequent radiative phase a dense shell forms and the expansion is driven by $p{\rm d}V$ work in this so-called pressure-driven snowplough phase ($r\propto t^{2/7}$, $v\propto t^{-5/7}$). In this phase the pressure in the dense shell is the same as in the shocked zone. When the pressure in the cavity has decreased enough, the remnant enters the momentum conserving phase ($r\propto t^{1/4}$, $v\propto t^{-3/4}$) in which the shell's momentum leads to further expansion of the bubble. We will now show, which power laws we found in the simulations. 
\subsection{Simulated pressure driven expansion}

In this phase the pressure inside the bubble pushes the shell into the ambient medium. Near the contact discontinuity a density peak forms. Behind the shock, at the outer side of the bubble's shell, a layer of heated, swept-up medium at $4$-times the ambient density develops\footnote{The maximal compression of an adiabatic mono-atomic gas leads to a factor $4$ in density.}. Despite radiative cooling losses the pressure in the shell gets much larger than the bubble pressure. Material starts to flow into the cavity and the bubble shell's density profile becomes symmetric. The largest cooling losses arise near the highest density gradient at the interface between the dilute bubble material and the swept-up ambient medium. For better resolved simulations the density in the cooling region becomes larger, but at the same time the cooling region becomes smaller. The simulations converge because for all of them the same amount of gas is compressed and cooled to the minimal temperature in the cooling table.

Since the swept-up shell is several cells wide, the retained kinetic energies converge as soon as the cell with the highest density has cooled to its equilibrium temperature and the cooling losses are dominated by the newly swept-up, compressed medium. The maximum luminosity is reached earlier for simulations with larger cells, since lower resolution will mix more of the hot gas in the bubble with the swept-up medium and thus enhance the cooling losses.

In models that take the wind of the progenitor of the SN into account, the maximal luminosity has two peaks and occurs later, since in a wind bubble the blast wave suffers no radiative losses until the wind shell is hit. In this case the bouncing SN blast wave inside the wind cavity causes double peaks in the loss rate: The first maximum in the loss is reached when the cavity wall is compressed and kinetic energy is converted to thermal energy and the second peak is found when the wall expands and thermal energy is converted back to kinetic energy. Due to the reflection of this wave inside the cavity the interaction of the wave and the cavity wall causes periodic conversions between thermal and kinetic energy with decreasing peak loss values until the SN wave is damped away.

During phases in which the pressure of the adiabatic expansion of the hot dilute (and therefore not cooling) interior of the bubble pushes the shell \citep[c.f.][]{Ostriker1988RvMP60p1,McKee1977ApJ218p148}, the change of momentum
\begin{eqnarray}
 \frac{4\pi \rho}{3} \frac{{\rm d} \left(\left(r\left(t\right)\right)^3 \frac{{\rm d} r\left(t\right)}{{\rm d} t}\right)}{{\rm d} t} = \underbrace{4\pi \left(r\left(t\right)\right)^2}_{\rm bubble\,surface}p_{\rm bubble}
\end{eqnarray}
can be combined with the law of adiabatic expansion 
\begin{eqnarray}
 \frac{p_{\rm bubble}\left(t\right)}{p_{\rm bubble}\left(0\right)} = \left( \frac{ r\left(t\right)}{ r\left(0\right)}\right)^{-3\gamma}
\end{eqnarray}
and an adiabatic exponent of $\gamma=\frac{5}{3}$. This way the exponents of $r$ become
\begin{eqnarray}
 3N+(N-1)-1&=&(2-5)N \nonumber \\
 N &=& 2/7 \nonumber
\end{eqnarray}
and thus dimensional analysis leads to
\begin{eqnarray}
 r\left(t\right) &=& c t^{2/7} \nonumber \\
 r\left(t\right) &=& \sqrt[7]{\frac{147r_0^5p_0}{2\rho}} t^{2/7} \label{eq:prs:r}
\end{eqnarray}
(c.f. eq.~12 of \citet{McKee1977ApJ218p148} for the pressure-driven phase: $r\left(t\right)=10^{-0.32}\sqrt[7]{\frac{r_c^2E_{\rm SN}}{n_0}} t^{2/7}$), which in turn leads to a velocity of
\begin{eqnarray}
 \frac{{\rm d} r\left(t\right)}{{\rm d} t} = \frac{2}{7}\sqrt[7]{\frac{147r_0^5p_0}{2\rho}} t^{-5/7} \label{eq:prs:v} 
\end{eqnarray}
and a kinetic energy of
\begin{eqnarray}
 E_{\rm kin} &=& \frac{mv^2}{2} = c t^{-4/7} \label{eq:prs:ek} \\
 {\rm with} && c= \frac{8\pi\rho}{147}\sqrt[7]{\frac{147r_0^5p_0}{2\rho}} \qquad .\nonumber
\end{eqnarray}

The best fits to the $40$~K models for times between the time of maximal luminosity $t_0$ and the time when the pressure inside the bubble has decreased to the ambient pressure (Table~\ref{tab:NOwind:PdV}) is, however, $E\propto t^{-0.7}$, $r\propto t^{0.272}$, $v\propto t^{-0.75}$. These fits rather resemble the behaviour of the momentum-conserving phase. Our models show that the pressure inside the bubble is much lower than the pressure in the shell. Thus, the overpressure of the shell also drives the bubble-expansion into the ambient medium. Table~\ref{tab:NOwind:PdV:shell} lists the times, when the shell pressure becomes larger than the bubble pressure. These times mark the end of the pressure driven phase and very close to these times (near $8$~kyr) a ``knee'' can be seen in Fig.~\ref{fig:NOwind:Ek+t}. Moreover, the best fits for the radius and the velocity in this short period of time is in agreement with fits of a pressure driven phase. The total kinetic energy decreases more slowly than a pressure driven fit would predict, since not all the kinetic energy is stored in the shell.

\citet{TenorioTagle1990MNRAS244p563} and \citet{TenorioTagle1996AJ111p1641} report hot swept-up matter separating the CD several parsecs from the outer shock for their SN explosion in a homogeneous medium. This is also seen in our simulation with $n_0=1$~\mcm, $T=100$~K. The CD and the outward shock are at the same radius as reported by \citet{TenorioTagle1990MNRAS244p563}. In our simulations the hot material between the CD and the thin dense shell (with a sub-parsec shell width, created by a sound wave from the reverse shock) is hot shocked swept-up ISM.
\subsection{Simulated momentum conservation}

Comparing the pressure inside the bubble to the pressure of the ambient medium shows that at $13\,t_0$ the $T=1\,000$~K model is already in the momentum conserving phase whereas the $40$~K model is still pressure driven. 
The times when the pressure inside the bubble has decreased to the ambient pressure are listed in Table~\ref{tab:NOwind:PdV}.
\begin{table}
\begin{tabular}{ccrc}
\multicolumn{1}{c}{$p$ [erg \mcm]} &$\Delta x$  [pc]        & \multicolumn{1}{c}{$t$ [kyr]}      &  $E_{\rm kin}$ [$10^{49}$~erg] \\
\hline \\[-1.0em]
 $3.99 \times 10^{-11}$ & $0.032$ & $34.5$ & $6.32$ \\
 $1.83 \times 10^{-12}$ & $0.032$ & $118.5$ & $2.58$\\
 $1.83 \times 10^{-12}$ & $0.016$ & $147.0$ & $2.12$\\
 $1.83 \times 10^{-12}$ & $0.008$ & $174.0$ & $1.85$\\
\end{tabular}
\caption{Times when the pressure inside the bubble has decreased to the ambient pressure in a model without prior winds. The SN is placed in a homogeneous ambient medium with a density of $2.2\times 10^{-22}$~g~\mcm. The ambient medium is in cooling-heating equilibrium.}
\label{tab:NOwind:PdV}
\end{table}
\begin{table}
\begin{tabular}{ccrc}
\multicolumn{1}{c}{$p$ [erg \mcm]} & $\Delta x$ [pc]      & \multicolumn{1}{c}{$t$ [kyr]} & \multicolumn{1}{c}{$t$ [kyr]}  \\
                                 &             & peak & average  \\
\hline \\[-1.0em]
 $3.99 \times 10^{-11}$ & $0.032$ & $6.5$ & $7.5$ \\
 $1.83 \times 10^{-12}$ & $0.032$ & $9.5$ & $9.5$\\
 $1.83 \times 10^{-12}$ & $0.016$ & $8$   & $8$\\
 $1.83 \times 10^{-12}$ & $0.008$ & $6.5$ & $6.5$\\
\end{tabular}
\caption{Times when the peak pressure in the shell becomes larger than the pressure inside the bubble in a model without prior winds. The SN is placed in a homogeneous ambient medium with a density of $2.2\times 10^{-22}$~g~\mcm. The ambient medium is in cooling-heating equilibrium.}
\label{tab:NOwind:PdV:shell}
\end{table}
Assuming that all ambient medium is swept-up in a thin, dense shell, this shell is at radius $r\left(t\right)$ moving with a velocity of $\frac{{\rm d} r(t)}{{\rm d} t}$ at time $t$.
 Momentum conservation
\begin{eqnarray}
 \frac{4\pi\rho}{3} \frac{{\rm d} \left(r\left(r\left(t\right)\right)^3 \frac{{\rm d} r\left(t\right)}{{\rm d} t}\right)}{{\rm d} t} = 0
\end{eqnarray}
leads to a radius of
\begin{eqnarray} 
 r\left(t\right) = b \sqrt[4]{a + 4t} \label{eq:mom:r}
\end{eqnarray}
and a velocity of
\begin{eqnarray}
 \frac{{\rm d} r\left(t\right)}{{\rm d} t} = b \left(a+4t\right)^{-0.75} \qquad, \label{eq:mom:v} 
\end{eqnarray}
 which leads to a kinetic energy of
\begin{eqnarray}
 E_{\rm kin} &=& \frac{mv^2}{2} = c \left(a+4t\right)^{-0.75} \label{eq:mom:ek} \\
 {\rm with} && c=\frac{2\pi\rho b^5}{3} \nonumber
\end{eqnarray}
where $a$, $b$ and $c$ are constants. This function was used to fit the kinetic energy evolution of models after the times listed in Table~\ref{tab:NOwind:PdV}.
\begin{figure}
 \includegraphics[width=\columnwidth, trim = 2mm  1mm 2mm 1mm, clip]{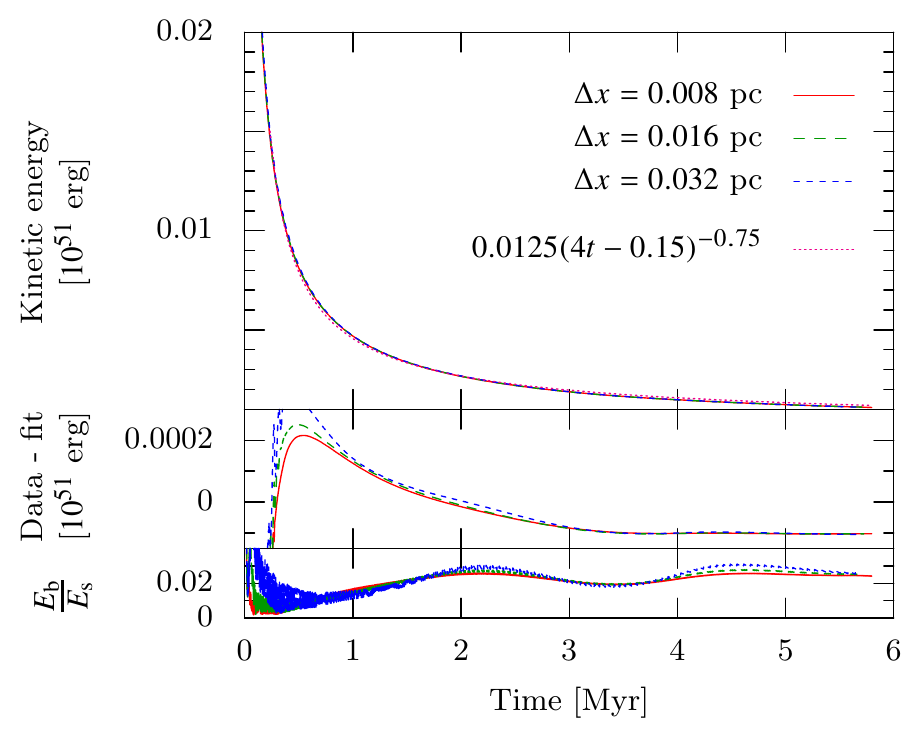}
\caption{Fit of a momentum conserving shell to the data. The middle panel shows the deviations from the fit. It can be seen that the kinetic energy decays more slowly than a momentum conserving model predicts. This indicates that the widening of the over-pressured shell contributes to the growth of the cavity. In the lowest panel the kinetic energy of bubble-gas is compared to the kinetic energy of the dense shell. The oscillations are caused by a wave travelling inside the cavity (see text).}
\label{fig:NOwind:Ek+t+fit}
\end{figure}
The fits of the bubble radius, the shell velocity and the kinetic energy showed that the kinetic energy decreases more slowly than eq.~\ref{eq:mom:ek} predicts (resp. the shell moves faster). The ratio between the shell's kinetic energy to the bubble's kinetic energy and the deviations of the fit from the kinetic energy (Fig.~\ref{fig:NOwind:Ek+t+fit}) show that the overpressure in the cavity wall leads to an expansion of the shell into the cavity and a high pressure wave starts to run back and forth in the cavity and the impacts on to the shell increase the shell velocity.

The best fit to the bubble radius after the end of the pressure driven phase is $r\propto t^{0.28}$. Thus, this fit for the radius of the simulated bubbles is rather a pressure-driven fit (eq.~\ref{eq:prs:r}) than a momentum-conserving fit (eq.~\ref{eq:mom:r}). The velocity $v\propto t^{-0.77}$ and kinetic energy $E\propto t^{-0.78}$, however, cannot be fitted with the pressure driven model.

The time when the shell velocity reaches the sound speed can be estimated from the fits by setting eq.~\ref{eq:mom:v} equal to the sound speed. All fits predicted a shorter time and a higher kinetic energy than the simulation data.
\section{Convergence}\label{sect:convergence}
\subsection{Temporal resolution}
In our simulations the time-step is limited by the CFL condition, which ensures that gas cannot travel more than a cell length per time-step. Thus, we can reduce the time-step via reducing the cell size $\left(\frac{\Delta x}{2}\right)$ or via reducing the factor in the CFL condition $\left(\frac{{\rm CFL}}{2}\right)$. I.e. the time-step for a simulation with CFL=$0.3$ is similar to the time-step in a simulation with CFL=$0.6$ and twice the number of cells per parsec. The time-steps of these two simulations differ a little, since variations in the velocities caused by the spatial resolution are a second order effect on the time-step size. The maximum velocities at a given time in the different simulations vary by less than $10$ per cent. The location of the cell, which limits the time-step depends on the evolution of the model: after $1$~Myr the gas velocity in the outermost cell of the free streaming wind region limits the time-step whereas after $4$~Myr the sound speed in the shocked wind region near the bubble wall limits the time-step size.

The two shock Riemann solver's efficiency is independent of the time-step size (varied via the CFL and by changing the time-marching algorithm from second order Runge-Kutta to third order Runge-Kutta) whereas an approximate Riemann solver (``Roe'' in {\sc Pluto}, see Appendix Sect.~\ref{sect:solver}) gets more efficient for larger time-steps, since the energy loss at the reverse shock occurs less often.
\subsection{Riemann solver and spatial interpolation scheme}\label{sect:solver}
We compared simulations in which the Riemann problem at each cell interface was solved in different ways:  the HLLC method \citep{HLLCsolver} only takes the fastest leftwards and rightwards moving characteristic into account and solves the problem with two intermediate states separated by a CD. It is thus very efficient, but also leads to the most diffusive solution of the three methods we compare. The Roe solver \citep{RoeSolver, RoeSlope} provides an exact solution of a linearized Jacobian. I.e.\ it keeps all $7$ characteristics, but treats all of them as simple waves. It is a shock capturing scheme, and can resolve a CD in approximately $3$ grid cells. Its known downsides are that it sometimes creates unphysical fluxes, that it can lead to negative thermal energies since it conserves total energy, and that it can create expansion shocks instead of expansion waves. In our simulation the Roe solver led to energy losses at the slowly moving reverse shock. This can be seen as damped oscillations in the shocked wind. The two shock solver solves the problem iteratively. It is a piecewise parabolic method: The states left and right of the interface are assumed to be parabolic (and not constant). As a consequence it allows for steepening near discontinuities. For more insight on the hydro solver, we refer the reader to \citet{Mignone2007ApJS170p228,Mignone2012ApJS198p7}.

In the simulations\footnote{This is a different set from the simulations in Table~\ref{tab:newGrid}.} with initial densities of $\rho=2.2\times 10^{-22}$~g~\mcm, pressures of $p=1.48\times 10^{-12}$~erg~\mcm, resolutions of $\Delta x=0.032$~pc and extreme mass loss ($500$~\Msol, which is much too high, but was used for tests of the kinetic energy fraction) in the SN, the two shock solver\citep{twoShock} ($1.8\times 10^{49}$~erg when the shell speed reaches the sound speed) is more efficient than the Roe \citep{RoeSolver, RoeSlope} solver ($1.5\times 10^{49}$~erg) and less efficient than the HLLC solver \citep{HLLCsolver} ($2.2\times 10^{49}$~erg). This is the expected behaviour, since the HLLC solver is the most diffusive of the three solvers and hence the density and temperature gradients at the contact discontinuity are shallower and thus the temperature in the first cell, which is dense enough to cool is smaller than in simulations with the two shock solver. On the other hand the Roe solver has the aforementioned problems with energy losses at the reverse shock. 

Actually all solvers produce oscillations inside the shocked wind region. A test with a constant wind showed that these oscillations are not caused by changes of the wind power (since they are also observed in a simulation with a constant wind).

To avoid energy losses at the reverse shock, the spatial interpolation scheme should allow for large gradients in this region. The {\sc Pluto} \citep{Mignone2007ApJS170p228} code's ``WENO3'' scheme \citep{WENO3} leads to a weighted essentially non oscillatory reconstruction of the primitive variables which reaches third order accuracy. It is suited for smooth data and led to a lower efficiency and stronger oscillations in the shocked wind region than the ``LINEAR'' scheme that carries out a piecewise total variation diminishing linear interpolation leading to second order accuracy in space. Also ``WENO3'' produces a density drop\footnote{Rather a dent or ``negative spike'' than a drop -- just one cell has a lower value.} on the inside of the shell, which leads to code crashes.
\section{Comparison to Tenorio Tagle et al. 1990}\label{sect:moreplots}
\begin{figure}
\includegraphics[width=\columnwidth, trim = 2mm  1mm 1mm 1mm]{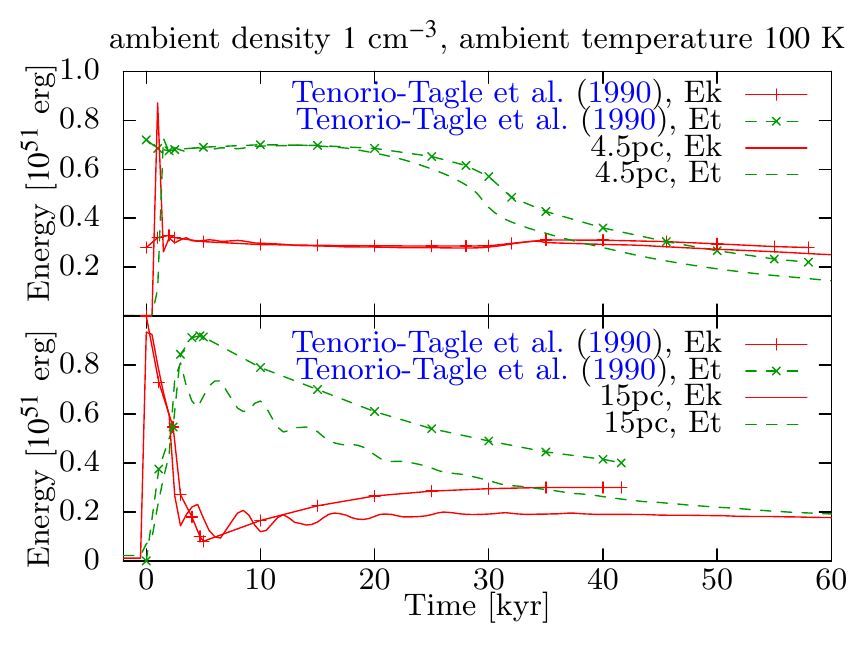} 
\caption{Comparison to Fig.~11c in \citet{TenorioTagle1990MNRAS244p563}}
\label{fig:compTenorio}
\end{figure}
Fig.~\ref{fig:compTenorio} compares the energy content of our simulations with constant winds producing a $4.5$~pc or $15$~pc cavity to Fig.~11c in \citet{TenorioTagle1990MNRAS244p563}.

\bsp	
\label{lastpage}
\end{document}